\documentclass[letterpaper,11pt]{article}

\pdfoutput=1

\PassOptionsToPackage{nosort}{cite}

\usepackage{amssymb}
\usepackage{amsfonts}
\usepackage{dsfont}
\usepackage{mathrsfs}
\usepackage{enumitem}
\usepackage[T1]{fontenc}
\usepackage[british]{chet}
\usepackage[font=small,format=hang,labelfont={sf,bf}]{caption}
\usepackage{float}
\usepackage{tikz}
\usepackage{pgfplots}

\usetikzlibrary{decorations.pathreplacing, arrows.meta, decorations.markings, bending, calc}

\usepgfplotslibrary{colorbrewer}
\pgfplotsset{width=14cm,compat=1.9}

\newcommand{\lsp}{\hspace{1pt}}

\newcommand{\lnsp}{\hspace{-1pt}}

\renewcommand{\geq}{\geqslant}

\allowdisplaybreaks

\definecolor{forestgreen}{rgb}{0.0, 0.27, 0.13}
\definecolor{darkblue}{rgb}{0.0, 0.0, 0.55}
\definecolor{darkred}{rgb}{0.55, 0.0, 0.0}
\hypersetup{colorlinks,
           linkcolor={forestgreen},
           citecolor={darkblue},
           urlcolor={darkred},
           pdftitle={Redundancy Channels in the Conformal Bootstrap},
           pdfdisplaydoctitle
}

\title{Redundancy Channels in the Conformal Bootstrap}

\author{Stefanos R.\ Kousvos$^{a,b}$ and Andreas Stergiou$^c$}

\affiliation{$^a$Department of Physics, University of Pisa and INFN, section of Pisa,\\[-1pt] Largo Pontecorvo 3, I-56127 Pisa, Italy\\[5pt]
$^b$Department of Physics, University of Torino and INFN, section of Torino,\\[-1pt] Via P.\ Giuria 1, 10125 Torino, Italy\\[5pt]
$^c$Department of Mathematics, King's College London, Strand, London WC2R 2LS, United Kingdom}

\abstract{A method for obstructing symmetry enhancement in numerical conformal bootstrap calculations is proposed. Symmetry enhancement refers to situations where bootstrap studies initialised with a certain symmetry end up allowing theories with higher symmetry. In such cases, it is shown that redundant operators in the less symmetric theory can descend from primary scaling operators of the more symmetric one, motivating the imposition of spectral gaps that are justified in the former but not the latter. The same mechanism can also be used to differentiate between decoupled and fully coupled theories which otherwise have the same global symmetry. A systematic understanding of this mechanism is developed and applied to distinguish the cubic from the $O(3)$ model in three dimensions, where a strip of disallowed parameter space, referred to as the cubic redundancy channel, emerges once a gap associated with a redundant operator of the cubic theory is imposed. The channel corresponds precisely to the region of parameter space where the assumed cubic symmetry would be enhanced to $O(3)$.}

\date{July 2025}

\begin{document}

\maketitle

\toc

\section{Introduction}
The numerical conformal bootstrap \cite{Rattazzi:2008pe, Poland:2018epd} has given us unprecedented access to strongly-coupled conformal physics. Most notably, it has revolutionised our understanding of the operator spectrum of the 3D Ising model \cite{El-Showk:2012cjh, El-Showk:2014dwa, Kos:2014bka, Kos:2016ysd, Chang:2024whx}, as well as that of $O(N)$ models \cite{Kos:2013tga, Kos:2015mba, Chester:2019ifh, Chester:2020iyt}. While these successes fully attest to the power and importance of the numerical conformal bootstrap, it is of great value to assess its strength in less symmetric cases, with richer operator spectra. This programme has been hindered by the problem of accidental enhancement of global symmetry, i.e.\ the problem of bootstrap constraints obtained with some global symmetry in mind allowing a theory with larger global symmetry \cite{Poland:2011ey, Stergiou:2018gjj, Li:2020bnb, Kravchuk:2021akc}. In this work, we provide a methodology to overcome this problem, and demonstrate explicitly how it works for cubic global symmetry in $d=3$ dimensions. A large part of the perturbative spectrum of this theory is already known from $\varepsilon$ expansion methods \cite{Antipin:2019vdg, Bednyakov:2023lfj,Henriksson:2025hwi,Henriksson:2025vyi}, further motivating a non-perturbative bootstrap study.

Our proposal is centred around redundant operators. In a Lagrangian formalism, redundant operators are proportional to the equation of motion and can be removed by field redefinitions. As such, they do not generate genuine physical effects and are modded out of the CFT spectrum. This, however, does not mean that they can be ignored in treatments of renormalisation \cite{Brown:1979pq, Brown:1980qq} or in discrete spin systems \cite{PhysRevB.32.5851} and Monte Carlo simulations. Indeed, they are known to be crucial for proper accounting of operator mixing and the derivation of anomalous dimensions, as well as the satisfaction of Ward identities. The fact that redundant operators vanish on-shell implies that their correlation functions are contact terms (semi-local or ultra-local). Such correlation functions are of no relevance to the numerical conformal bootstrap, which is based on the constraints of unitarity and crossing symmetry of position-space four-point functions in configurations where all operator insertion points are separated.

Despite this, redundant operators have in fact been used in the numerical conformal bootstrap for a long time, albeit through their absence as exchanged operators in relevant operator product expansions (OPEs). For example, the 3D Ising island is obtained assuming a single relevant scalar $\mathbb{Z}_2$-odd operator, commonly denoted by $\sigma$, exchanged in the $\sigma\times\epsilon$ OPE, where $\epsilon$ is the leading scalar $\mathbb{Z}_2$-even operator (beyond the identity). In a weakly coupled description, the next-to-leading scalar $\mathbb{Z}_2$-odd operator, $\sigma'$, would be given schematically by $\phi^3$, but, as we will review below, this operator becomes redundant. This makes $\sigma'$ schematically of the form of $\phi^5$ and motivates a large gap between $\sigma$ and $\sigma'$, which ends up being crucial in isolating the 3D Ising model in a bootstrap island. Such perturbatively inspired considerations were given a more formal flavour in \cite{Rychkov:2015naa}.

The main message of this work is that gaps associated with redundant operators are much more widely useful in the numerical conformal bootstrap. For our case of interest, that of unwanted symmetry enhancements from $H$ to $G$ with $H$ a subgroup of $G$, the crucial observation is that redundant operators in the theory with symmetry $H$ may be drawn from scaling, non-redundant operators of the theory with symmetry $G$. This motivates the introduction of gaps in the $H$-symmetric theory that are absent in the corresponding sector of the $G$-symmetric one, if there are no nearby non-redundant operators with the same quantum numbers in the $H$-symmetric theory. Such gaps, then, preclude the enhancement of $H$ to $G$. Equation of motion inspired gaps also allow us to differentiate between decoupled and fully coupled theories, based on the observation that a fully coupled theory has one stress-energy energy tensor, whereas decoupled theories have multiple.

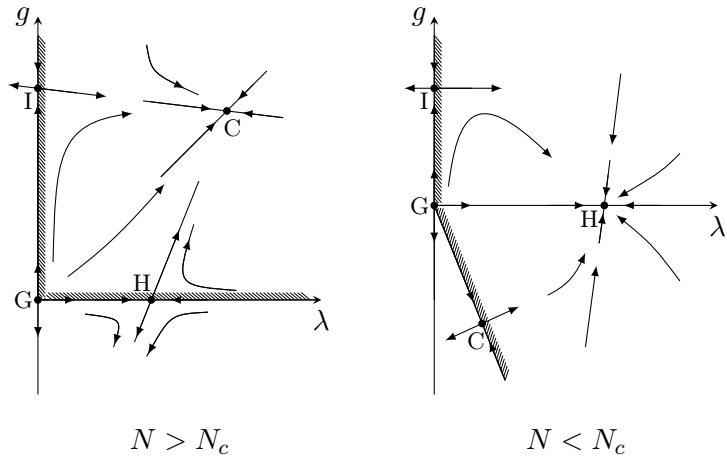
\begin{figure}[H]
  \centering
  \begin{tikzpicture}[scale=1.25]
    \draw[decoration={border,segment length=0.5mm,amplitude=1.5mm,angle=40},
      postaction={decorate,draw,ultra thin}] (0,2.8)--(0,0);
    \filldraw[fill=white, color=white] (0,0.02) rectangle (0.5,-0.5);
    \draw[decoration={border,segment length=0.5mm,amplitude=1.5mm,angle=140},
      postaction={decorate,draw,ultra thin}] (0.2,0)--(2.95,0);
    \draw[decoration={border,segment length=0.5mm,amplitude=0.35mm,angle=140},
      postaction={decorate,draw,ultra thin}] (0.05,0)--(0.12,0);
    \draw[decoration={border,segment length=0.5mm,amplitude=1mm,angle=140},
      postaction={decorate,draw,ultra thin}] (0.15,0)--(0.2,0);
    \draw[->,>=stealth] (0,0)--(3,0) node[below] {$\lambda$};
    \draw[->,>=stealth] (0,0)--(0,3) node[left=-1pt] {$g$};
    \draw (0,0)--(0,-1);
      \filldraw (0,0) circle (1pt) node[left=-2pt]
        {\footnotesize{G}}; 
      \filldraw (1.2,0) circle (1pt)
        node[label={[label distance=-6pt, xshift=2pt]160:{\footnotesize{H}}}] {}; 
      \filldraw (2,2) circle (1pt)
      node[label={[shift={(0.07,-0.6)}]
        {$\text{\footnotesize{C}}$}}] {}; 
      \filldraw (0,2.24) circle (1pt)
        node[label={[shift={(-0.13,-0.55)}]
        {$\text{\footnotesize{I}}$}}]{}; 
    %
      \draw[-latex] (0,0)--(0.4,0); 
      \draw[-latex] (0,0)--(0,0.4); 
      \draw[-latex] (0,0)--(0,-0.4); 
      \draw[-latex, shorten <=0.5cm, shorten >=1.5cm] (0,0) .. controls
      (0.9,0.7) .. (2,2);
      \draw[-latex, shorten <=0.5cm, shorten >=1.5cm] (0.15,0) .. controls
      (0.2,2) .. (2,2);
      \draw (0.5,-0.15) .. controls (0.9,-0.2) .. (0.8,-0.5);
      \draw[-latex, shorten >=0.03cm] (0.5,-0.15) .. controls (0.9,-0.2) ..
        (0.8,-0.5);
    %
      \draw[-latex] (1.8,0)--(1.4,0); 
      \draw[-latex] (0.6,0)--(1,0); 
      \draw[-latex,shorten >=1.69cm,shorten <=-0.6cm] (1.2,0)--(2,2); 
      \draw[shorten >=1cm] (1.2,0)--(2,2);
      \draw[-latex] (1.2,0)--(1.04,-0.4); 
      \draw[-latex, shorten >=0.2cm] (2.1,0.1) .. controls (1.45,0.15) ..
        (1.65,0.8);
      \draw (2.1,0.1) .. controls (1.45,0.15) .. (1.65,0.8);
      \draw (1.8,-0.13) .. controls (1.4,-0.15) .. (1.15,-0.6);
      \draw[-latex, shorten >=0.03cm] (1.8,-0.13) .. controls (1.4,-0.15) ..
        (1.15,-0.6);
    %
      \draw[-latex, shorten <=2.3cm, shorten >=0.2cm] (0,0)--(2,2);
      \draw[shorten <=2.3cm, shorten >=-0.5cm] (0,0)--(2,2);
      \draw ($(2,2)!-0.6cm!(0,2.24)$)--(2,2);
      \draw[-latex, shorten >=0.2cm] ($(2,2)!-0.6cm!(0,2.24)$)--(2,2);
      \draw[-latex, shorten >=0.2cm] ($(2,2)!-0.6cm!(0,0)$)--(2,2);
      \draw (1.15,2.7) .. controls (1.2, 2.4) .. (1.7,2.17);
      \draw[-latex, shorten >=0.2cm] (1.15,2.7) .. controls (1.2, 2.4)
        .. (1.7,2.17);
      \node at (1.5,-1.5) {$N>N_c$};
      \draw[-latex, shorten >= 1.6cm] (0,2.24)--(2,2);
      \draw[shorten <= 1.4cm] (0,2.24)--(2,2);
      \draw[-latex, shorten >=0.2cm, shorten <= 1.5cm] (0,2.24)--(2,2);
      \draw[-latex, shorten >= 0.2cm] (0,0)--(0,2.24);
      \draw[-latex, shorten >= 0.2cm] (0,2.44)--(0,2.24);
      \draw[-latex, shorten <= 2cm, shorten >= -0.4cm] (2,2)--(0,2.24);
  \end{tikzpicture}
  \hspace{0.5cm}
    \begin{tikzpicture}[scale=1.25]
    \filldraw[fill=white, color=white] (0,1) rectangle (0.1,0.93);
    \draw[decoration={border,segment length=0.5mm,amplitude=1.5mm,angle=320},
      postaction={decorate,draw,ultra thin}] (0.75,-0.85)--(0,1);
    \filldraw[fill=white, color=white] (0,1.05) rectangle (0.1,0.98);
    \draw[decoration={border,segment length=0.5mm,amplitude=1.5mm,angle=40},
    postaction={decorate,draw,ultra thin}] (0,2.8)--(0,1.02);
    \filldraw[fill=white, color=white] (0,1.02) rectangle (0.1,0.98);
    \draw (0.75,-0.85)--(0,1);
    \draw[->,>=stealth] (0,1)--(3,1) node[below] {$\lambda$};
    \draw[->,>=stealth] (0,0)--(0,3) node[left=-1pt] {$g$};
    \draw (0,0)--(0,-1);
      \filldraw (0,1) circle (1pt) node[left=-2pt]
        {\footnotesize{G}}; 
      \filldraw (1.8,1) circle (1pt)
        node[label={[shift={(-0.2,-0.56)}]{\footnotesize{H}}}] {}; 
      \filldraw (0,2.24) circle (1pt)
        node[label={[shift={(-0.13,-0.55)}]
        {$\text{\footnotesize{I}}$}}]{}; 
      \filldraw (0.507,-0.25) circle (1pt)
        node[label={[shift={(-0.08,-0.6)}]{\footnotesize{C}}}] {};
    %
      \draw[-latex] (0,1)--(0.4,1); 
      \draw[-latex] (0,1)--(0,1.4); 
      \draw[-latex] (0,1)--(0,0.6); 
      \draw[-latex, shorten >=0.3cm] (0,1)--(0.507,-0.25);
      \draw[-latex] (0.15,1.2) .. controls (0.25,2) and (0.45,2.3) ..
      (1.25,1.5);
    %
      \draw[-latex] (2.4,1)--(2,1); 
      \draw[-latex] (1.2,1)--(1.6,1); 
      \draw[-latex, shorten >=0.1cm, shorten <=0.4cm] (1.9,1.8)--(1.8,1);
      \draw[latex-, shorten >=-0.75cm, shorten <=0.8cm] (1.8,1)--(1.9,1.8);
      \draw[shorten <=0.6cm] (1.9,1.8)--(1.8,1);
      \draw[-latex, shorten >=0.1cm] (1.745,0.6)--(1.8,1);
      \draw (1.745,0.6)--(1.8,1);
      \draw[-latex, shorten >=0.2cm] (2.6,0.2) .. controls (2.3,0.6) ..
        (1.8,1);
      \draw[-latex, shorten >=0.2cm] (2.6,1.55) .. controls (2.4,1.35) ..
        (1.8,1);
      \draw[-latex, shorten >=0.8cm] (1.6,-0.5)--(1.8,1);
    %
      \draw[-latex, shorten >=0.25cm] (0.75,-0.85)--(0.507,-0.25); 
      \draw[-latex] (0.507,-0.25)--(0.9,-0.07);
      \begin{scope}[rotate around={180:(0.507,-0.25)}]
        \draw[-latex] (0.507,-0.25)--(0.9,-0.07);
      \end{scope}
      \draw[-latex] (1.2,0.05) .. controls (1.4,0.15) and (1.5,0.25) .. (1.6,0.5);
    %
      \draw[-latex, shorten >= 1.6cm] (0,2.24)--(2,2.24);
      \draw[-latex, shorten >= 0.2cm] (0,0)--(0,2.24);
      \draw[-latex, shorten >= 0.2cm] (0,2.44)--(0,2.24);
      \draw[-latex, shorten <= 2cm, shorten >= -0.4cm] (2,2.24)--(0,2.24);
    \node at (1.5,-1.5) {$N<N_c$};
  \end{tikzpicture}
  \caption{Depending on the value of $N$, the IR stable fixed point is either the $O(N)$ model, denoted by H, or the hypercubic theory, denoted by C. The free theory is denoted by G and decoupled Ising models by I. The region between the hatched lines is the basin of attraction of the IR stable fixed point.}
  \label{fig:flows}
\end{figure}

While our proposal enjoys a high-degree of generality, which we will outline in various examples arising via the $\varepsilon$ expansion \cite{Pelissetto:2000ek, Osborn:2017ucf, Rychkov:2018vya}, we will mostly examine it here in the context of a theory with cubic global symmetry in $d=3$. A theory with hypercubic global symmetry, $C_N=\mathbb{Z}_2{\!}^N\rtimes S_N$, can be defined in $d=4-\varepsilon$ starting with $N$ scalar fields, $\phi_i$, $i=1,\ldots,N$, with interactions described by the action
\begin{equation}\label{eq:action}
    S=\int d^{\lsp 4-\varepsilon}x\,\big(\tfrac12\partial^\mu\phi_i\lsp\partial_\mu\phi_i+\tfrac{1}{8}\lambda(\phi^2)^2+\tfrac{1}{4!}g\lsp\delta_{ijkl}\phi_i\phi_j\phi_k\phi_l\big)\,,
\end{equation}
where $\delta_{ijkl}$ is the generalised Kronecker delta symbol, which is equal to one when all indices take the same value and zero otherwise. Besides the free theory ($\lambda=g=0$) and the theory of $N$ decoupled Ising models ($\lambda=0, g\ne0$), this action is known to have two fully interacting fixed points, namely the $O(N)$ model ($\lambda\ne0, g=0$) and the hypercubic model ($\lambda g\ne0$). At large $N$, it is known that the infrared (IR) stable fixed point is the hypercubic one, while, as $N$ is lowered, the hypercubic fixed point approaches the $O(N)$ one. At some critical value of $N$, $N_c$, these two fixed points pass through each other. At that point, they exchange their stability properties and below $N_c$ it is the $O(N)$ theory that is IR stable \cite{PhysRevB.8.3323, PhysRevB.8.4270, D_J_Wallace_1973, Pelissetto:2000ek, Osborn:2017ucf}. This is summarised in Fig.\ \ref{fig:flows}. 

The value $N_c$ is equal to four at leading order in the $\varepsilon$ expansion, but gets corrected order by order in $\varepsilon$. Early studies can be found in \cite{Mudrov:1998hi, Varnashev:1999ze, Carmona:1999rm, Folk_2000}, and more recent work includes \cite{Adzhemyan:2019gvv, Chester:2020iyt, Binder:2021vep, Aharony:2022ajv, Hasenbusch:2022zur}. Notably, a six-loop analysis in the $\varepsilon$ expansion suggested that $N_c<3$, so that for $N=3$ the IR stable fixed point is the cubic one \cite{Adzhemyan:2019gvv}. A bootstrap analysis confirmed this in a fully non-perturbative way directly in $d=3$, by showing that in the $O(3)$ model the rank-four traceless symmetric tensor, from which the symmetry-breaking cubic deformation of the $O(3)$ model is drawn, is relevant \cite{Chester:2020iyt}. A state of the art Monte Carlo study also corroborated the statements above, further showing that the next-to-leading scalar singlet in the cubic theory is irrelevant \cite{Hasenbusch:2022zur}. Notably, the critical value of flavours for which the $O(N)$ and $C_N$ fixed points exchange stability is very close to $N=3$, evident by the fact that the leading four-index traceless symmetric scalar operator of the $O(3)$ fixed point has a dimension very close to $\Delta = 3$. This operator can be explicitly added to the $O(3)$ Lagrangian, triggering a flow to the $C_3$ fixed point. Given that this operator is almost marginal, the flow itself is ``short''. This fact can be exploited in conformal perturbation theory \cite{Rong:2023owx}. The shortness of the flow leads to near degeneracy of dimensions of operators in these theories. This further complicates the act of telling these theories apart in the bootstrap, providing an additional motivation for using the $C_3$ theory as a toy model.

It would be desirable to isolate the cubic theory in an island for high-precision study with the numerical conformal bootstrap. This pursuit provided the initial motivation for bootstrap studies of theories with hypercubic symmetry \cite{Rong:2017cow, Stergiou:2018gjj, Kousvos:2018rhl, Kousvos:2019hgc}. However, this effort ran into the problem of accidental enhancement to $O(3)$ global symmetry. In this work we are able to isolate the cubic theory in an island that does not include the $O(3)$ model. While our island is large, we believe that the road to producing a small bootstrap island and performing a detailed spectrum analysis of the cubic theory in $d=3$ is now open.

In the next section we describe in detail essential aspects of redundant operators. In Section \ref{sec:redboot} we describe how gaps inspired by our perturbative understanding of redundant operators can be useful in various bootstrap setups. In Section \ref{sec:redapps} we identify systems of correlation functions that need to be considered in the minimal setting to enable the use of gaps associated with redundant operators. Our numerical results for the cubic theory are also discussed in Section \ref{sec:redapps}. We conclude in Section \ref{sec:conc}, while some details of our numerical implementation are included in Appendix \ref{AppendixParameters}.

\section{Redundant operators}
\subsection{The equation of motion}
The canonical, free, massless scalar theory is described by the action
\begin{equation}
    S_{\text{free}}=\int d^{\lsp d}x\,\tfrac12\partial^\mu\phi\lsp\partial_\mu\phi\,.
\end{equation}
The equation of motion of $\phi$ is
\begin{equation}\label{eq:eom_free}
    \partial^2\phi=0\,.
\end{equation}
This equation implies that correlation functions involving the $\partial^2\phi$ operator will be pure contact terms, with no support when all insertions are taken to lie at non-coincident points. It is clear that one can assign a scaling dimension $\Delta_\phi+2=\frac12(d+2)$ to $\partial^2\phi$, which is reflected in the two-point function
\begin{equation}
    \langle\partial^2\phi(x)\lsp\partial^2\phi(0)\rangle\propto\partial^2\delta^{(d)}(x)\,.
\end{equation}

Let us now consider a quartic interaction term for $\phi$ and work in $d=4-\varepsilon$ dimensions. The action becomes
\begin{equation}\label{eq:Sint_single}
    S_{\text{int}}=\int d^{\lsp 4-\varepsilon}x\,\big(\tfrac12\partial^\mu\phi\lsp\partial_\mu\phi+\tfrac{1}{4!}\lambda_0\phi^4\big)\,,
\end{equation}
where $\lambda_0$ is the bare coupling, and the equation of motion of $\phi$ is
\begin{equation}\label{eq:eom_int}
    \partial^2\phi-\tfrac{1}{3!}\lambda_0\phi^3=0\,.
\end{equation}
We would like to elucidate here the implications of \eqref{eq:eom_int} for the operator spectrum of the CFT that is obtained at the Wilson--Fisher fixed point of this theory at $\lambda_\ast=\frac{16\pi^2}{3}\varepsilon$, where the renormalised coupling $\lambda$ is defined by $\lambda+L(\lambda)=\lambda_0\mu^{-\varepsilon}$, where $L(\lambda)$ is a series expansion in $1/\varepsilon$, starting at order $1/\varepsilon$. To do this, we need to be careful about renormalisation effects. In particular, there is non-trivial operator mixing between $\phi^3$ and $\partial^2\phi$ as soon as $\lambda\neq0$.

For $\lambda=0$, $\phi^3$ is a scaling operator and $\partial^2\phi$ a redundant operator of the free fixed point. Their scaling dimensions are $\Delta_{\phi^3}=3\Delta_\phi=\frac32(d-2)$ and $\Delta_{\partial^2\phi}=\Delta_\phi+2=\frac12(d+2)$, respectively. Of course, $\partial^2\phi$ is modded out of the free CFT spectrum. When $\lambda\ne0$, $\phi^3$ mixes with $\partial^2\phi$. Since $\partial^2\phi$, whose renormalisation is determined by that of $\phi$, does not mix with $\phi^3$, the corresponding $2\times 2$ mixing matrix will be upper triangular:
\begin{equation}
 \begin{pmatrix}\label{mixingPhi} [\tfrac{1}{3!}\lambda\mu^\varepsilon\phi^3] \\
 [\partial^2\phi]\end{pmatrix}
 =
  \begin{pmatrix}
   Z_\phi^{1/2}  & Z_\phi^{-1/2}-Z_\phi^{1/2} \\
   0 & Z_\phi^{-1/2}
  \end{pmatrix}
  \begin{pmatrix} \tfrac{1}{3!}\lambda_0\phi^3 \\
  \partial^2 \phi
  \end{pmatrix},
\end{equation}
where we use square brackets to denote renormalised operators. It is known that the renormalisation of $\phi^3$ is fixed by $Z_\phi$ \cite{Brezin:1974zr}. (One can also prove this and find the form of \eqref{mixingPhi} using diagrammatic arguments.) The left eigenvectors of the mixing matrix in \eqref{mixingPhi} give the scaling operators at the IR fixed point:
\begin{equation}
 \begin{pmatrix}\label{mixingPhi2} E_\phi \\
 [\partial^2\phi]\end{pmatrix}
 =
  \begin{pmatrix}
   -1 & 1 \\
   0 & 1 
  \end{pmatrix}
 \begin{pmatrix} [\tfrac{1}{3!}\lambda\mu^\varepsilon\phi^3] \\
 [\partial^2\phi]\end{pmatrix},
\end{equation}
with corresponding eigenvalues $Z_\phi^{1/2}$ for $E_\phi$ and $Z_\phi^{-1/2}$ for $[\partial^2\phi]$.

From here one we will be more sloppy and write expressions like
\begin{equation}
 \begin{pmatrix}\label{mixingPhi3} E_\phi \\
 \partial^2\phi\end{pmatrix}
 =
  \begin{pmatrix}
   *  & * \\
   0 & * 
  \end{pmatrix}
 \begin{pmatrix} \phi^3 \\
 \partial^2\phi\end{pmatrix}
\end{equation}
to capture the operator mixing. The upshot of our analysis is that the proper scaling operators in the interacting fixed point defined through the renormalisation group (RG) are $\partial^2\phi$ and the equation of motion operator $E_\phi$. $E_\phi$ is however removed from the CFT, which contains only the operators that have non-zero correlation functions at non-coincident points, while $\partial^2\phi$ remains. In the rest of the present manuscript we will write equations like \eqref{mixingPhi3} to express the scaling operators in terms of building blocks made out of powers of the fundamental operator and insertions of derivatives.

In the free fixed point, $\phi^3$ is a genuine scaling operator, while $\partial^2\phi$ is redundant. In the interacting fixed point, $\partial^2\phi$ is a genuine scaling operator, while $\phi^3$ combines with $\partial^2\phi$ to form the redundant operator $E_\phi$. This point of view provides a precise, renormalised perturbation theory interpretation of \cite{Rychkov:2015naa}. Standard renormalisation logic does not play well with the interpretation that the equation of motion sets two operators equal, in this case $\phi^3$ and $\partial^2\phi$. In particular, one may not think of $\phi^3$ as a scaling operator of the interacting fixed point in renormalised perturbation theory. Therefore, in the  Wilson--Fisher theory, \cite[Eq.\ (2.10)]{Rychkov:2015naa} only makes sense as $\Delta_{\partial^2\phi}=\Delta_\phi+2$. Correlation functions of $[\frac{1}{3!}\lambda\mu^{\varepsilon}\phi^3]$ differ from correlation functions of $[\partial^2\phi]$ by contact terms, which are crucial for properly defining correlation functions as distributions.

Let us also include here a quick argument that shows that the dimension of the equation of motion operator is given by $d-\Delta_\phi$, regardless of fixed point. Denote the equation of motion operator by $E_\phi$, which is equal to $\partial^2\phi$ in the free theory and $[\partial^2\phi]-[\frac{1}{3!}\lambda_*\mu^{\varepsilon}\phi^3]$ in the interacting one. Then, consider the correlation function
\begin{equation}
    \begin{aligned}
        \langle E_\phi(x)\phi(x_1)\cdots\phi(x_n)\rangle&=-\Big{\langle} \frac{\delta S}{\delta\phi(x)}\phi(x_1)\cdots\phi(x_n)\Big{\rangle}\\
        &=\sum_{r=1}^n\delta^{(d)}(x-x_r)\langle\phi(x_1)\cdots\phi(x_{r-1})\phi(x_{r+1})\cdots\phi(x_n))\rangle\,.
    \end{aligned}
\end{equation}
Clearly,
\begin{equation}
    \Delta_{E_\phi}=d-\Delta_\phi\,.
\end{equation}
This is valid in both UV and IR fixed points (for the proper corresponding $E_\phi$).

Multiplet recombination can be recast as follows. When interactions cause the multiplet of $\phi$ to no longer be truncated, there must be some other multiplet that gets removed to account for the apparent mismatch in the number of states between the ultraviolet (UV) and IR fixed points. Indeed, the operator $\phi^3$ of the free theory combines with $\partial^2\phi$ as discussed above, in precisely the right way for the resulting renormalised operator of the interacting theory to be removed by the equation of motion of the interacting theory. From a more formal, algebraic point of view, redundant operators are viewed as null states that decouple from all correlators of the CFT, and the number of such states is preserved for CFTs connected by an RG flow. This provides a non-Lagrangian understanding of redundant operators.

\subsection{A note on shift symmetry breaking}
As discussed, going from the free theory to the interacting fixed point we lose the primary operator $\phi^3$ due to equation of motion effects. This can also be viewed from a symmetry breaking point of view, which will be a central part of our analysis in the rest of the paper. Note that the free scalar theory has a shift symmetry $\phi \rightarrow \phi+c$, where $c$ is a constant. It is clear that $\partial^2 \phi$ is invariant under such shifts, but $\phi^3$ is not and hence in the free theory these two operators cannot mix. In the interacting theory, however, shift symmetry is broken and $\phi^3$ can now mix with $\partial^2\phi$, which removes a primary from the spectrum of the interacting theory. As we will see below, this also happens when breaking other symmetries, such as breaking $O(N)$ to $C_N$. However, the primary we will lose will not be of $\phi^3$ type. In fact, we will see in specific examples that we can lose $\phi^4$ and $\partial \phi^4$-type primaries, among others. What type of primary is lost depends on the specific example considered.

\subsection{Composite operators involving the equation of motion operator}
The discussion above shows that any operator that is zero by the equation of motion will result in a non-trivial renormalisation constraint. This includes composite operators formed as a product of some operator with $E_\phi$.

A first application of this observation comes about when considering higher-spin currents of the free theory. These operators, denoted by $J_{\mu_1\ldots\mu_{\ell}}, \ell>2$, are no longer conserved in the interacting theory, and so the conformal multiplets of $J_{\mu_1\ldots\mu_{\ell}}$ are not short. However, there exist primary operators of the free theory of spin $\ell-1$ and free-theory dimension $2\Delta_\phi+\ell+1=d+\ell-1$ \cite{Skvortsov:2015pea} that mix with $\partial^{\mu_1}J_{\mu_1\ldots\mu_{\ell}}$ in the proper way to yield redundant operators in the Wilson--Fisher theory. The case $\ell=2$, corresponding to the stress-energy tensor, $T_{\mu\nu}$, is excluded from this discussion. Indeed, there is no primary, spin-one operator of dimension $d+1$ in the free theory with which $\partial^\mu T_{\mu\nu}$ can mix \cite{Rychkov:2015naa, Skvortsov:2015pea}. As a result, $\partial^\mu T_{\mu\nu}=0$ must continue to hold in the Wilson--Fisher theory, after use of the equation of motion. This was checked in \cite{Brown:1979pq}, where it was shown that $\partial^\mu T_{\mu\nu}$ is proportional to the redundant operator $\partial_\nu\phi \lsp E_\phi$.

An operator like $\phi E_\phi$ can also be seen to be finite and thus easily promoted to a renormalised operator with scaling dimension $d$ \cite{Brown:1979pq}. Such redundant operators are of great interest in our work, but to be more specific we need to consider a generalisation of \eqref{eq:Sint_single} to
\begin{equation}\label{eq:Sint_multi}
    S_{\text{int}}=\int d^{\lsp 4-\varepsilon}x\,\big(\tfrac12\partial^\mu\phi_i\lsp\partial_\mu\phi_i+\tfrac{1}{4!}\lambda_{ijkl}\phi_i\phi_j\phi_k\phi_l\big)\,,
\end{equation}
where $\lambda_{ijkl}$ is a symmetric tensor. The equation of motion now is
\begin{equation}
    (E_\phi)_i=0\,,\qquad (E_\phi)_i=\partial^2\phi_i-\tfrac{1}{3!}\lambda_{ijkl}\phi_j\phi_k\phi_l\,.
\end{equation}
If we assume that \eqref{eq:Sint_multi} has a global symmetry $G$ under which $\phi_i$ transforms in the vector representation, then the product $\phi_i(E_\phi)_j$ needs to be decomposed into irreducible representations of $G$. In the $O(N)$ model, for example, this would give a singlet, a rank-two traceless symmetric and an antisymmetric redundant operator. The singlet redundant operator in $\phi_i(E_\phi)_j$ would be proportional to the trace of the stress-energy tensor $T^\mu{\!}_\mu$, and the antisymmetric one to the divergence of the global symmetry current. In the hypercubic theory, there would be four redundant operators, for the rank-two traceless symmetric representation of the $O(N)$ model is reducible under the hypercubic group.

\subsection{Examples where the equation of motion does not lead to a gap}\label{nonremoval}
While in the present work we will focus on gaps that can be imposed when primary operators are removed from the spectrum due to equation of motion effects, it is useful to also understand when and why this might not happen. An obvious example is already found by considering a theory with two quartic in $\phi$ singlets, like the hypercubic theory \eqref{eq:action}. The equation of motion will remove one combination of $\partial^2\phi_i$ with the two $\phi^3$-type operators that arise from the potential of \eqref{eq:action}, but another combination will remain.

Another simple example is encountered when considering the rank-two traceless symmetric representation of $O(N)$. In this case, at engineering dimension four, we have three building blocks, namely $\phi_i \phi_j \phi^2$, $\partial^\mu\phi_i\lsp \partial_\mu \phi_j$ and $\partial^2 t_{ij}$, where we have omitted subtractions of traces for simplicity (and we have also omitted the building block $\phi_i \partial^2 \phi_j + \phi_j \partial^2 \phi_i$ since it is not linearly independent). We denote the leading rank-two traceless symmetric operator, of engineering dimension two, by $t_{ij}$. Their mixing matrix will be, schematically,
\begin{equation}
 \begin{pmatrix}\label{mixingtij} t'_{ij}\\  \phi_{(i} (E_\phi)_{j)}  \\
 \partial^2 t_{ij}\end{pmatrix}
 =
  \begin{pmatrix}
   \ast & \ast  & \ast \\
   \ast & \ast  & \ast \\
   0 & 0 & \ast 
  \end{pmatrix}
  \begin{pmatrix} \phi_i \phi_j \phi^2 \\  \partial^\mu\phi_i\lsp \partial_\mu \phi_j \\
  \partial^2 t_{ij}
  \end{pmatrix},
\end{equation}
where subtractions of traces have again been dropped. The two zeroes on the last line of the mixing matrix are due to the fact that $\partial^2 t_{ij}$ is renormalised once $t_{ij}$ is, i.e.\ there is no mixing with $\phi_i \phi_j \phi^2$ and $\partial^\mu\phi_i\lsp \partial_\mu \phi_j$ necessary to render it a finite operator. This is however not true for $\phi_i \phi_j \phi^2$ and $\partial^\mu\phi_i\lsp \partial_\mu \phi_j$ themselves, which are not automatically renormalised scaling eigenoperators. The operators in the left-hand side of \eqref{mixingtij} are $t'_{ij}$ which is a primary, $\phi_{(i} (E_\phi)_{j)} $ which is proportional to the equation of motion, and $\partial^2 t_{ij}$ which is the descendant of $t_{ij}$.

Due to the building block $\partial^\mu\phi_i\lsp \partial_\mu \phi_j$, then, we still have a left over primary after use of the equation of motion and therefore no special gap due to the equation of motion can be imposed in the corresponding bootstrap problem. In the examples below, where we consider the non-conservation of currents, the building block $\partial^\mu\phi_i\lsp \partial_\mu \phi_j$ will be eliminated due to the antisymmetry of the representation in which the broken current will transform.

\section{Redundant operators relevant to the conformal bootstrap}\label{sec:redboot}
We now proceed to outline the general methodology for locating the precise sectors of a given theory in which redundant operators will generate large gaps in the spectrum. This relies on identifying composite operators involving the equation of motion that make specific parts of the spectrum in a given theory sparse. To be a bit more specific, in the hypercubic case, which is our central example, we will show that carefully studying the breaking of the global symmetry current as $O(N) \rightarrow C_N$ will lead us to a large gap that can efficiently segregate the $O(N)$ from the $C_N$ theory in numerical studies. We will show that carefully studying the reduction of the number of conserved stress-energy tensors when going from decoupled Ising models to hypercubic theories, we can find gaps that strongly exclude the decoupled Ising model from parameter space. Putting these two results together, we can exclude two of the three fixed points of the hypercubic action \eqref{eq:action}.\footnote{There is of course also the free theory, but in the numerical conformal bootstrap this is always essentially trivially excluded.} This leaves the hypercubic theories alone in parameter space, modulo other theories which we refer to as hidden sectors that will be discussed at the end of this section.

\subsection{Broken currents}
Before proceeding to specific examples which will elucidate all necessary technical details, let us outline the general behaviour of broken currents when breaking a continuous group\footnote{Or a group with a continuous part, such as e.g.\ $O(m)^n\rtimes S_n$.} $G$ to a subgroup $H<G$, where at least part of the continuous symmetry is broken. In particular, within the numerical conformal bootstrap examples we will discuss, $G$ will more often than not be $O(N)$. The group $G$ will have one or more conserved currents satisfying $\partial_\mu J^\mu = 0$. The conservation of the current can be seen as a direct consequence of the equation of motion. For example, if $G=O(N)$ we have
\begin{equation}
    \partial_\mu J^\mu_{ij}\sim \phi_i \partial^2 \phi_j - \phi_j \partial^2 \phi_i= \phi_i (E_\phi)_j - \phi_j (E_\phi)_i=0\,,
\end{equation}
since $(E_\phi)_i=0$.

When a continuous symmetry (or part of it) is broken, the conservation of the current (or some components of it) is obviously violated. In the context of perturbative field theory, this happens through operator mixing, which as we explained earlier is very closely related to the concept of multiplet recombination. In particular, under the subgroup $H$ the divergence of the broken current (or of specific broken components) will now be in the same global symmetry representation as some operator $\mathcal{O}$ that is close to marginality. These operators will mix to create a scaling eigenoperator of the type
\begin{equation}
    \mathcal{O}_{\text{scaling}}=\partial_\mu J^\mu - \mathcal{O} \sim E_\phi =0\,.
\end{equation}
This is usually written in the literature as $\partial_\mu J^\mu = \mathcal{O}$, but from a perturbative perspective the only scaling eigenoperator that exists containing $\mathcal{O}$ is $\partial_\mu J^\mu -\mathcal{O}$, i.e.\ $\mathcal{O}$ on its own is not an eigenstate of the dilatation operator, as can e.g.\ be checked in the $\varepsilon$ expansion.\footnote{Similarly to $\phi^3$ by itself not being a scaling operator at the interacting Ising fixed point.}

The exercise that now needs to be carried out is the determination of the global symmetry representation that the broken currents transform in under in $H$. Subsequently, we must enumerate the building blocks available to build operators, and see if there are enough of them to build a primary once the broken current ``eats'' the equivalent(s) of $\mathcal{O}$ above. If there are not enough building blocks left to build a primary, this will create a large gap in the spectrum of $H$ that did not exist in $G$. To make the discussion more precise, we proceed to explicit examples computable in the $d=4-\varepsilon$ expansion concerning theories that have been of interest to the numerical conformal bootstrap.

\subsubsection{Hypercubic theory}
In this example we have $G=O(N)$ and $H=C_N$. Here the conservation of all components of the conserved current $J^\mu_{ij}$ breaks. Let us set $N=3$ for simplicity. The components of the current can be written out as 
\begin{gather}
J^{\mu}_{ij}
 =
  \begin{pmatrix} \phi_1\partial^\mu \phi_2 - \phi_2 \partial^\mu \phi_1 \\  \phi_2\partial^\mu \phi_3 - \phi_3 \partial^\mu \phi_2 \\ \phi_3\partial^\mu \phi_1 - \phi_1 \partial^\mu \phi_3 \end{pmatrix}.
\end{gather}
It is clear that each element above is antisymmetric under swapping index values;\footnote{In hypercubic theories representations are determined by their transformation properties under permutations of index values (e.g.\ $1 \leftrightarrow 2$), whereas in $O(N)$ representations are determined by the properties under permutations of indices (e.g.\ $i \leftrightarrow j$).} it is thus an operator that transforms in the antisymmetric rank-two irrep of the hypercubic group, which we call $B$, see \cite[Table 1]{Bednyakov:2023lfj}. For the general group theory of replica groups and how to build operators in them see also \cite{Kousvos:2021rar,Kousvos:2024dlz}.

The divergence of this current will of course be of the form 
\begin{equation}
    \partial_\mu J^{\mu}_{ij}=\phi_i \partial^2 \phi_j - \phi_j \partial^2 \phi_i\,,
\end{equation}
and we explicitly see that the divergence of the broken current will transform under the $B$ irrep, at Lorentz spin zero, and engineering dimension four in $d=4$.  Whatever mixes with this operator must have the exact same properties. The only available candidate is 
\begin{equation}
    B_{ij}=\phi_i \phi_j^3-\phi_j\phi_i^3\,.
\end{equation}
These two operators will thus mix, and the mixing matrix at the hypercubic fixed point will be of the form
\begin{equation}
 \begin{pmatrix}\label{currentmixingcubic} (\phi E_\phi)_{ij} \\  \partial_\mu J^\mu_{ij} \end{pmatrix}
 =
  \begin{pmatrix}
   \ast & \ast \\
   0 & \ast 
  \end{pmatrix}
  \begin{pmatrix} \phi_i \phi_j^3 -\phi_j \phi_i^3 \\  \phi_i \partial^2 \phi_j - \phi_j \partial^2 \phi_i \end{pmatrix},
\end{equation}
where the operators on the left-hand side are now renormalised scaling eigenoperators at the hypercubic fixed point. The first of the two eigenoperators is a composite equation of motion operator, specifically
\begin{equation}
    (\phi E_\phi)_{ij}= \phi_i (E_\phi)_j - \phi_j (E_\phi)_i =\partial_\mu J^\mu_{ij}-c B_{ij}=0\,,
\end{equation}
where $c$ is some coefficient that won't be important. On the other hand $\partial_\mu J^\mu_{ij}$ is simply the descendant of $J^\mu_{ij}$, which is now an ordinary primary operator (i.e.\ it is no longer short). We thus observe that our two building blocks for constructing operators get consumed creating a redundant operator and a descendant, and there is no freedom left to construct a primary!

At the $O(N)$ fixed point, $B_{ij}$ and $\phi_i \partial^2 \phi_j -\phi_j \partial^2 \phi_i$ do not mix since they belong to different irreps of $O(N)$, namely the rank-four traceless symmetric and rank-two antisymmetric, respectively. In this case, $B_{ij}$ is now part of a primary, and $\partial_\mu J^\mu$ is a composite equation of motion, i.e.\ redundant, as explained earlier. 

This will be the precise criterion for telling these theories apart in the conformal bootstrap. That is, the first $B$ spin-zero operator will be a $\phi^6$ operator at the hypercubic fixed point, since the only $\phi^4$ candidate gets eaten by the divergence of the broken current, whereas if our crossing equations are satisfied by the $O(N)$ fixed point, the leading $B$ operator will be of $\phi^4$ type, and thus of much lower dimension. For the perturbative determinations of these operators see \cite{Bednyakov:2023lfj,Henriksson:2025hwi,Henriksson:2025vyi} and \cite{Bednyakov:2021ojn,Henriksson:2022rnm,Bednyakov:2023lfj} respectively, we give the Padé resummations in Table \ref{TableB}. 

\begin{table}[H]
\centering
\begin{tabular}{|c|c|lr|}
\hline
 Irrep & $C_3$ & $O(3)$ & 
\\\hline
$B$   & 5.301 & 2.992 & $(t_4)$ 
\\\hline
\end{tabular}
\caption{Scaling dimension of the leading spin-zero $B$ operator at the cubic and $O(3)$ fixed points. At the cubic fixed point, it is a $\phi^6$-type operator whose dimension is available to order $\varepsilon^5$ \cite{Henriksson:2025hwi,Henriksson:2025vyi}, and we performed a Padé$_{2,3}$ resummation. At the $O(3)$ fixed point, it is a $\phi^4$ type operator in the rank-four traceless symmetric irrep of $O(3)$, whose dimension is known to order $\varepsilon^6$ \cite{Bednyakov:2021ojn,Henriksson:2022rnm,Bednyakov:2023lfj}, and we perform a Padé$_{3,3}$ resummation. The dimension of this operator at the $O(3)$ fixed point has also been determined via Monte Carlo methods to be 2.987(4) \cite{Hasenbusch:2011zwv}, while the conformal bootstrap has bounded its dimension to not be above 2.99056 \cite{Chester:2020iyt}.}\label{TableB}
\end{table}

\subsubsection{Hypertetrahedral theory}
In this example we have $G=O(N)$ and $H=T_N=S_{N+1}\times\mathbb{Z}_2$. Such models were first bootstrapped in \cite{Rong:2017cow, Stergiou:2018gjj}. For $N=3$, the tetrahedral and cubic theories are identical in the $\varepsilon$ expansion. For $N>5$ there are two distinct, fully-interacting theories with $T_N$ symmetry, one of which is IR stable. These coincide at leading order in $\varepsilon$ when $N=5$. For $N=4$ there is only one hypertetrahedral theory at leading order in the $\varepsilon$ expansion. Just like the hypercubic case, all components of the $O(N)$ global symmetry current are broken in the corresponding hypertetrahedral theories.

To realise this symmetry on $N$ scalar fields, it is convenient to define $N+1$ vectors in $N$-space, $(e_N)_i^\alpha$, $i=1,\ldots,N$, $\alpha=1,\ldots,N+1$, which give the locations of the $N+1$ vertices of an $N$-dimensional hypertetrahedron. Starting from $N=1$ with $(e_1)_1^1=-(e_1)_1^2=-\frac{1}{\sqrt{2}}$, we define, recursively,
\begin{equation}
\begin{aligned}
    (e_N)_i^\alpha &= (e_{N-1})_i^\alpha\,, \qquad i=1,\ldots,N-1,\;\alpha=1,\ldots,N\,,\\
    (e_N)_N^\alpha&=-\sqrt{\frac{1}{N(N+1)}}\,,\qquad \alpha=1,\ldots, N\,,\\
    (e_N)_i^{N+1}&=\sqrt{\frac{N}{N+1}}\delta_i{\hspace{-0.4pt}}^N\,.
\end{aligned}
\end{equation}
These vectors satisfy
\begin{equation}
    \sum_{\alpha} (e_N)_i^\alpha=0\,,\qquad
    \sum_{\alpha}(e_N)_i^\alpha(e_N)_j^\alpha=\delta_{ij}\,,\qquad
    (e_N)_i^\alpha(e_N)_i^\beta=\delta^{\alpha\beta}-\frac{1}{N+1}\,.
    \label{eq:evectorrules}
\end{equation}
Starting with $\phi_i$ we may define $\phi^\alpha=(e_N)_i^\alpha\phi_i$, such that $\sum_{\alpha}\phi^\alpha=0$ due to the first of \eqref{eq:evectorrules}.

The operator that mixes with the divergence of the $O(N)$ current is given by
\begin{equation}
    V^{\alpha\beta}=(\phi^\alpha)^3\phi^\beta-(\phi^\beta)^3\phi^\alpha+\frac{1}{N+1}\phi^3(\phi^\alpha-\phi^\beta)\,,
\end{equation}
where $\phi^3=\sum_\alpha(\phi^\alpha)^3$. This operator was first given in \cite[Appendix D]{Osborn:2017ucf} and descends from the rank-four traceless symmetric tensor of the $O(N)$ theory. A combination of this operator with the divergence of the $O(N)$ current is removed in the hypertetrahedral theory via the corresponding equation of motion, leaving the now non-zero divergence of the $O(N)$ current as a descendant in the spectrum. This is similar to what happens in the hypercubic models.

\subsubsection{MN model}
We proceed to our next example, which concerns the family of CFTs obtained by considering $n$ coupled replicas of the $O(m)$ model. These are referred to as MN models, and within the bootstrap have been studied in \cite{Stergiou:2019dcv, Henriksson:2021lwn, Kousvos:2021rar}. Their global symmetry group is $M\!N_{m,n}=O(m)^n\rtimes S_n$, which is a subgroup of $O(mn)$. They can be seen a generalisation of the hypercubic groups, and also as special cases of replica groups $K^n\rtimes S_n$, where $K$ is some arbitrary group. For a bootstrap friendly treatment of the group theory of replica groups see \cite{Kousvos:2021rar, Kousvos:2024dlz}.

In these theories, it is convenient to write the fundamental field as a matrix, $\phi^a_i$. The upper index transforms under $S_n$ and the lower under $O(m)$. While most details will follow through in the same way as the preceding hypercubic group, the conserved current of $O(mn)$ only partially breaks. This is because $M\!N_{m,n}$ has $n$ conserved currents, one for each factor of $O(m)$. Let us call the conserved current of each $O(m)$ factor $J^{(r)\lsp\mu}_{ij}$, where $r$ runs from $1$ to $n$. The matrix of all conserved currents,
\begin{gather}
A^{\mu}{}_{ij}^{ab}
 =
  \begin{pmatrix} J^{(1)\lsp\mu}_{ij} & 0 & \cdots & 0 \\ 0 & J^{(2)\lsp\mu}_{ij} & \cdots & 0\\ \vdots & \vdots & \ddots & \vdots \\ 0&0& \cdots & J^{(n)\lsp\mu}_{ij} \end{pmatrix},
\end{gather}
furnishes an irreducible representation (see \cite{Kousvos:2021rar,Kousvos:2024dlz} for group theory details) of $M\!N_{m,n}$ that we call $A$. These currents are all conserved, and thus we have
\begin{equation}
    \partial_\mu A^{\mu}{}_{ij}^{ab} =0\,. 
\end{equation}

The dimension of the representation $A$ is $n\frac{m(m-1)}{2}$, whereas the conserved current of the original $O(mn)$ symmetry had $\frac{mn(mn-1)}{2}$ components. This means that $\frac{n(n-1)m^2}{2}$ components of the original current get broken. Schematically, under $O(mn) \rightarrow O(m)^n\rtimes S_n$,
\begin{equation}
    J_\mu^{O(mn)} \rightarrow J_\mu^{\text{MN}}+J_\mu^{\text{broken}}\,.
\end{equation}
We now need to find out the irrep of $M\!N_{m,n}$ to which $J_\mu^{\text{broken}}$ belongs. It turns out to be the generalisation of the irrep $B$ of the hypercubic groups we saw earlier. The dimension of this irrep is precisely $\frac{n(n-1)m^2}{2}$. The generalisation of \eqref{currentmixingcubic} thus becomes
\begin{equation}
 \begin{pmatrix}\label{currentmixingMN} (\phi E_\phi)^{ab}_{ij} \\  {(\partial_\mu J^\mu)}^{ab}_{ij} \end{pmatrix}
 =
  \begin{pmatrix}
   \ast & \ast \\
   0 & \ast 
  \end{pmatrix}
  \begin{pmatrix} \phi^a_i \delta^{bcde} \phi^c_j \phi^d_k \phi^e_k - \phi^b_j \delta^{acde} \phi^c_i \phi^d_k \phi^e_k \\  \phi^a_i \partial^2 \phi^b_j - \phi^b_j \partial^2 \phi^a_i \end{pmatrix}.
\end{equation}

The same discussion as for hypercubic then follows: if our crossing equations are satisfied by a true MN theory, then the leading $B$ scalar operator will be of $\phi^6$ type. However, if the symmetry of our crossing solution has enhanced to $O(mn)$, the leading $B$ operator in our spectrum will be a $\phi^4$ operator. Thus, a large gap in the $B$ sector is again the prescribed way to forbid symmetry enhancement.

\subsubsection{Bifundamental theory}

We now proceed to the so-called bifundamental theories, which have been of particular interest to the conformal bootstrap \cite{Nakayama:2014lva,Nakayama:2014sba,Henriksson:2020fqi,Dowens:2020cua,Reehorst:2024vyq}. These are scalar field theories with global symmetry $O(m)\times O(n)/\mathbb{Z}_2$, and are expressed naturally using a fundamental field with two indices $\phi_{ar}$ in the Lagrangian description. The name bifundamental derives from the fact that the index $a$ transforms in the vector representation of $O(m)$, and $r$ transforms in the vector of $O(n)$. Representations of this group can thus be described as a pair of labels $(R_1,R_2)$, denoting transformation properties under $O(m)$ and $O(n)$, respectively. The modded out factor of $\mathbb{Z}_2$ will not be of importance for us presently, but we retain it to differentiate from theories with more than one mass term, such as the biconical theories with $O(m)\times O(n)$ symmetry that we will consider below.\footnote{The group theoretical significance of the modded out $\mathbb{Z}_2$ is that the $\mathbb{Z}_2$ action on $\phi_{ar}$ coming from $O(m)$ is indistinguishable from the one coming from $O(n)$.}

As in the MN example we saw above, only some components of the conserved current of $O(mn)$ are broken, i.e.\
\begin{equation}
    J_\mu^{O(mn)} \rightarrow {J_1}_\mu^{O(m)\times O(n)/\mathbb{Z}_2}+{J_2}_\mu^{O(m)\times O(n)/\mathbb{Z}_2}+J_\mu^{\text{broken}}\,.
\end{equation}
As always, our task is to find the representation that $J_\mu^{\text{broken}}$ transforms in under $O(m)\times O(n)/\mathbb{Z}_2$. To do this, it is convenient to spell out explicitly the unbroken currents, of which we have two (one for $O(m)$, one for $O(n)$). These are 
\begin{equation}
\label{omoncurrent1}
    {J_1}^\mu_{ab} = \delta_{rs} (\phi_{ar}\partial^\mu \phi_{bs}-\phi_{br}\partial^\mu \phi_{as})\,,
\end{equation}
which is a conserved current of $O(m)$, and 
\begin{equation}
\label{omoncurrent2}
    {J_2}^\mu_{rs} = \delta_{ab} (\phi_{ar}\partial^\mu \phi_{bs}-\phi_{as}\partial^\mu \phi_{br})\,,
\end{equation}
which is a conserved current of $O(n)$. These transform respectively in the $(A,S)$ and $(S,A)$ representations. In this notation, the conserved current of $O(mn)$ can be written as
\begin{equation}
\label{omoncurrent3}
     J^\mu_{ab,rs} = \phi_{ar}\partial^\mu \phi_{bs}- \phi_{bs}\partial^\mu \phi_{ar}\,.
\end{equation}

It now becomes clear how to decompose the current in \eqref{omoncurrent3}, since the currents in \eqref{omoncurrent1} and \eqref{omoncurrent2} are simply its traces with respect to $\delta_{rs} $ and $\delta_{ab}$ respectively:
\begin{equation}
\begin{split}
\label{currentbranching}
     J^\mu _{ab,rs} &= \phi_{ar}\partial^\mu \phi_{bs}- \phi_{bs}\partial^\mu \phi_{ar}\\&= (\phi_{ar}\partial^\mu \phi_{bs}-\phi_{br}\partial^\mu \phi_{as})+ (\phi_{br}\partial^\mu \phi_{as}-\phi_{bs}\partial^\mu \phi_{ar})\\
     &=(\phi_{ar}\partial^\mu \phi_{bs}-\phi_{br}\partial^\mu \phi_{as})+ (\phi_{br}\partial^\mu \phi_{as}-\phi_{bs}\partial^\mu \phi_{ar})\\
     &\quad \pm \frac{1}{n}\delta_{rs}(\phi_{at}\partial^\mu \phi_{bt}-\phi_{bt}\partial^\mu \phi_{at})\pm \frac{1}{m}\delta_{ab}(\phi_{cr}\partial^{\mu} \phi_{cs}-\phi_{cs} \partial^\mu \phi_{cr})\\
     &=(AT)^\mu_{ab,rs}+(T\!A)^\mu_{ab,rs}+\delta_{rs}{J_1}^\mu_{ab}+\delta_{ab}{J_2}^\mu_{rs}\,,
\end{split}
\end{equation}
where as implied by their name in \eqref{currentbranching} the broken currents transform in the $(A,T)$ and $(T,A)$ irreps, respectively. Indeed, the dimensions of $(T,A)$ and $(A,T)$ add up to the difference in components of the $O(mn)$ and $O(m)\times O(n)/\mathbb{Z}_2$ currents.\footnote{In other words, $\frac{m(m-1)}{2} \frac{(n-1)(n+2)}{2} +\frac{ n(n-1)}{2} \frac{(m-1)(m+2)}{2} -\left( \frac{mn(mn-1)}{2} - \frac{m(m-1)}{2} - \frac{n(n-1)}{2}   \right)=0$.} 

The mixing matrix that we need to consider involves the $(A,T)$ and $(T,A)$ representations, at Lorentz spin zero and engineering dimension four, as usual. We have
\begin{gather}
 \begin{pmatrix}\label{currentmixingOmn} (\phi E_\phi)^{ab}_{rs} \\  {(\partial_\mu J^\mu)}^{ab}_{rs} \end{pmatrix}
 =
  \begin{pmatrix}
   \ast & \ast \\
   0 & \ast 
  \end{pmatrix}
  \begin{pmatrix} \phi^{(a}_t \phi^{b)}_{[r} \phi^c_{s]}\phi^{c}_t\\  \phi^{(a}_{[r} \partial^2 \phi^{b)}_{s]}  \end{pmatrix},
\end{gather}
where we have dropped traces for simplicity. The mixing matrix \eqref{currentmixingOmn} refers explicitly to the $(A,T)$ representation. We omit the one for $(T,A)$, since it can be obtained by trivially switching labels. 

The lesson relevant for bootstrap studies is that there will be large gaps in the spectrum with the $(T,A)$ and $(A,T)$ sectors at spin zero, which will allow us to segregate $O(m)\times O(n)/\mathbb{Z}_2$ theories in parameter space from $O(mn)$ symmetric ones.

\subsubsection{Biconical model}
Let us now discuss an example where we don't obtain a large gap in the spectrum, even though we do lose a $\phi^4$-type primary. All our examples so far have had only one quadratic invariant (operator of the type $\phi^2$). Fixed points with more are also common in the $\varepsilon$ expansion \cite{Osborn:2020cnf}. A simple example is the so-called biconical family of theories with $O(m)\times O(n)$ global symmetry \cite{Nelson:1974xnq, Kosterlitz:1976zza, Calabrese:2002bm}. These can be obtained by adding a $O(m)$-invariant action to an $O(n)$-invariant one, and then coupling them by their corresponding quadratic invariants:
\begin{equation}
    S_{O(m)\times O(n)} = S_{O(m)}+S_{O(n)} + \int d^{\lsp 4-\varepsilon}x\, \tfrac14 g (\phi_a\phi_a)(\chi_i \chi_i)\,,
\end{equation}
where $\phi_a$ is the fundamental field of the $O(m)$ part and $\chi_i$ that of the $O(n)$ part. The indices $a$ and $i$ run from one to $m$ and $n$, respectively.

We will label representations of the $O(m)\times O(n)$ symmetry by doublets $(R_1,R_2)$, where $R_1$ denotes the representation under $O(m)$ and $R_2$ under $O(n)$. The $O(m)\times O(n)$ symmetric theory possesses two conserved currents, one for $O(m)$ and one for $O(n)$, similarly to the case of bifundamental theories discussed above. It is easy to see that the components of the original $O(m+n)$ symmetry current that will no longer be conserved will be of the form 
\begin{equation}
    (J_\mu^\text{broken})_{ai}= \phi_a \partial_\mu \chi_i - \chi_i \partial_\mu \phi_a\,.
\end{equation}
This is because the fundamental field of the original $O(m+n)$ theory is simply the column vector $(\phi_a,\chi_i)^T$. It becomes evident that the broken current transforms in the bivector representation $(V,V)$. Indeed, $(V,V)$ has dimension $mn$ and the difference between the dimensions of the original and the unbroken currents is $\frac{(m+n)(m+n-1)}{2}-\frac{m(m-1)}{2}-\frac{n(n-1)}{2}=mn$.

In these biconical models, we have two equations of motion, $(E_\phi)_a=0$ and $(E_\chi)_i=0$. However, the operators of engineering dimension four that mix are $\partial^2(\phi_a\chi_i), \phi_a\partial^2\chi_i,\chi_i\partial^2\phi_a, \phi_a\chi_i\phi^2$ and $\phi_a\chi_i\chi^2$. Their mixing will result in two descendants, $\partial^2 (\phi_a \chi_i)$ and $\partial^\mu ({J_\mu}^\text{broken})_{ai}$, and two redundant operators, but will also leave behind one primary. Therefore, a gap in the $(V,V)$ sector would not be justified in biconical models. We thus see that, similarly to the examples in Section \ref{nonremoval}, it is very important to accurately count all building blocks to assess if a gap can actually be imposed.

\subsection{Broken stress-energy tensors}
Having provided an understanding of how broken symmetries can lead to large gaps in spectra of theories through the non-conservation of symmetry currents, we now proceed to show how one can also obtain large gaps by reducing the number of conserved stress-energy tensors. This occurs while comparing a decoupled theory, which may have a number $n>1$ of conserved stress-energy tensors, $T_i^{\mu\nu}, i=1,\ldots,n$, to a coupled theory which may only have one, namely $T^{\mu \nu}_{\text{coupled}} =  T^{\mu \nu }_1 +  T^{\mu \nu }_2 + \cdots +  T^{\mu \nu }_n$.
Thus, the conservation of $n-1$ stress-energy tensors will be broken, and one must find the representation under which these transform in the coupled theory, to see if a large gap is indeed created.

\subsubsection{Hypercubic theory example}
Let us proceed to an explicit example. In a theory of $N$ decoupled Ising models we have $N$ conserved stress-energy tensors. When transitioning to a theory of $N$ coupled Ising models (i.e.\ the hypercubic theory), only their sum remains conserved. The non-conserved components will transform in the $X$ Lorentz spin two representation of the hypercubic group. The divergence of these operators, i.e.\ $\partial_\mu X^{\mu \nu}$ will transform in the $X$ Lorentz spin-one representation, at engineering dimension five. This gives the precise mixing matrix we must consider. Primaries at engineering dimension five and Lorentz spin one can schematically be built using operators of the form 
\begin{equation}
    X_\mu \sim C_{abcd\lsp \lsp}\phi_a \phi_b \phi_c \partial_\mu \phi_d\,,
\end{equation}
see \cite{Henriksson:2025hwi,Henriksson:2025vyi}. One should not totally symmetrise $C_{abcd}$, since this would make $X_\mu$ a total derivative. Instead, $C_{abcd}$ should just be symmetric under permutations of its first three indices.

We thus have to consider the mixing between $X^\mu$ and $\partial_\nu X^{\mu\nu}$. This will look like
\begin{gather}
 \begin{pmatrix}\label{stressmixing} {\partial^\mu \phi\lsp E_\phi} \\  \partial_\nu X^{\mu\nu} \end{pmatrix}
 =
  \begin{pmatrix}
   \ast & \ast \\
   0 & \ast 
  \end{pmatrix}
  \begin{pmatrix} X^\mu \\  \partial_\nu X^{\mu\nu} \end{pmatrix},
\end{gather}
where the operator $\partial^\mu \phi E_\phi$ has dimension $d+1$. Indeed, looking at \cite[Table 17]{Bednyakov:2023lfj}, we see that there exists no $\partial \phi^4$-type primary in the irrep $X$ at spin one. (Notice, however, that for the irrep $Z$ in Table $19$ of the same paper, which is also rank-two symmetric, there does exist a $\partial \phi^4$-type primary.) Therefore, the first primary in the $X$ Lorentz spin-one representation will be of the form $\partial \phi^6$. However, in the decoupled Ising theory $\partial_\nu X^{\mu\nu}=0$, and there will thus be no mixing with $X^\mu$. This means that the leading primary will be of $\partial \phi^4$ type. In conclusion, as far as numerics are concerned, a large gap in the $X$ spin-one sector should prove sufficient to remove the decoupled Ising models from parameter space without affecting the fully interacting hypercubic theory.

\subsection{Hidden sectors}
\label{HiddenSymmetries}
In our discussion so far, we have described how to use equations of motion to remove CFTs from our parameter space. In the explicit example of the hypercubic Lagrangian, we were able to exclude all its fixed points, except for the one we wanted to study, i.e.\ the fully coupled hypercubic one. However, in a generic bootstrap study, our parameter space can also include other fixed points, derived from other Lagrangians.\footnote{We will not comment on potential fixed points not derivable from Lagrangians.} One example is scalar-fermion fixed points. However, the lowest-dimension operator $\phi$ of such theories is typically of dimension higher than in a pure scalar field theory. This is expected to be the case for potential gauge theories too.

Another example concerns long range models; for some with hypercubic symmetry see \cite{Benedetti:2020rrq}. However, these theories can be easily excluded by demanding the existence of a stress-energy tensor, and then imposing a gap in the spectrum after it, which is precisely what we will do later on.

There remains one class of examples, that is theories with global symmetry $G_1 \times G_2$, where $G_1$ is the symmetry we are trying to study, e.g. $G_1=C_N$ for hypercubic, and $G_2$ is some ``hidden sector''.\footnote{Some examples of this type of fixed point within the $\varepsilon$ expansion can be seen in \cite{Liendo:2022bmv}.} One way to exclude these theories is by looking at the Lorentz spin one singlet sector. Here, the leading operator is of $\partial \phi^6$-type in the theory without a hidden sector, see \cite[Table 8]{Bednyakov:2023lfj}. However, if there is a hidden sector, then calling $\chi^2$ its leading scalar singlet, we can write down the operator $\chi^2 \phi_i \partial_\mu \phi_i$ in the singlet spin-one representation. This operator is of $\partial \phi^4$ type, and cannot be removed with equation of motion arguments.

In conclusion, hidden sectors can be excluded by imposing a large gap in the $S$ spin-one sector. Notice that a theory with a hidden sector should in principle be easy to tell apart from one without, since the hidden sector will generate at least one more scalar singlet that is quadratic in the field. That being said, if this additional scalar singlet happens to be weakly coupled, the numerics may not be able to easily exclude it. A large gap in the singlet spin-one sector should prove more robust.

\section{Applications to the numerical conformal bootstrap}\label{sec:redapps}
\subsection{Choosing the correct correlator system}
In the preceding sections, we have shown that large gaps can appear for a variety of reasons, e.g.\ broken currents or broken stress-energy tensors. We now outline how to choose specific systems of correlators that manifest these large gaps in the spectrum. We do this once again through explicit examples, through which we expect the reader will be able to intuit the general picture.

\subsubsection{Hypercubic theory}
In the hypercubic fixed point, we saw that a large gap is created in the $B$ irrep at Lorentz spin zero due to the breaking of $J^\mu_{ij}$ of $O(N)$. Similarly, we saw that another large gap is created in the $X$ irrep at Lorentz spin one, due to the reduction of the amount of stress-energy tensors compared to decoupled Ising models.

Note that we want to find the OPE which exchanges the operators in question, where the externals are as light as possible, since in practice this affects numerical studies. For $B$ the appropriate OPE is
\begin{equation}
    X \times Z \sim Z + B +XZ\,,
\end{equation}
where now the operators on the right-hand side appear for both spins, since the OPE is between non-identical operators. The lightest $X$ and $Z$ operators are $\phi^2$-type operators in the perturbative limit. We remind the reader that while $B$ is exchanged as an irrep also in the OPE of the lightest operator in the theory $\phi$,
\begin{equation}
    \phi \times \phi \sim S+X+Z+B\,,
\end{equation}
it is only exchanged at odd spins, since it is antisymmetric under swapping the indices of the two $\phi$'s on the left-hand side.

Bootstrap intuition tells us that it is always a good idea to include in our system of external operators the lightest operators in the theory. Indeed, as we will soon see numerically, including $X$ and $Z$ as externals is sufficient to preclude symmetry enhancement, but is not numerically strong enough to provide an island (at least for the choices of gaps and numerical strength we tried). However, including $\phi$, i.e.\ considering a mixed correlator system involving $\phi$, $X$ and $Z$ as external operators, we were able to obtain an island. In fact, in order to just obtain an island (which will however still include $O(3)$) the $\phi$ and $Z$ system of externals suffices.

Similarly, in order to make the $X$ spin-one gap appear in the bootstrap, one needs to consider the OPE
\begin{equation}
    X \times S \sim X\,,
\end{equation}
which exchanges $X$ at all spins. The operators on the left-hand side will again be of $\phi^2$ type. Thus, a correlator system with $\phi$, $X$, $Z$ and $S$ will provide an island and systematically exclude the $O(3)$ and decoupled Ising CFTs. The only remaining CFTs in this parameter space\footnote{That we know of perturbatively, at least.} are those with hidden sectors we discussed earlier. That is, theories where the symmetry group would be $C_3 \times K$, where all external operators we are considering are charged under the $C_3$ part but are singlets under $K$. As discussed in Section \ref{HiddenSymmetries}, these can be excluded by putting a gap in the $S$ spin-one sector, which would be exchanged in the $S \times S^\prime$ OPE, where $S^\prime$ is $\phi^4$-type. However, we expect this to be unnecessary, since the spectrum of a hidden sector theory will be considerably more dense, hence it should already be excluded once we start adding a few gaps. Despite this, the fact that the leading $S$ spin-one operator is of particularly large dimension \cite[Table 8]{Bednyakov:2023lfj}, could significantly shrink an island, albeit at the cost of including yet another external which would make the numerics even more demanding. 

\subsubsection{Hypertetrahedral theory}
The analysis for hypertetrahedral theories is very similar to the one for cubic theories. That is, one should consider the OPE between the two representations called $Y$ and $V$ in \cite{Stergiou:2018gjj}. These are the two symmetric rank-two representations exchanged in the $\phi \times \phi$ OPE. This will in turn exchange the two index antisymmetric representation which hosts the broken current.

\subsubsection{MN model}
For MN theories, the analysis follows through similarly to the hypercubic case. That is, a $\phi$-$X$-$Z$ correlator system is sufficient to exclude enhancement to $O(mn)$ symmetry, and the $\phi$-$X$-$S$ system will exclude decoupled $O(m)$ models from parameter space. A system including all four as externals will exclude both $O(mn)$ and decoupled $O(m)$ models, leaving the MN theory on its own to study.

\subsubsection{Bifundamental theory}
In the case of bifundamental theories, we wish to make the $AT$ and $TA$ representations appear in the OPEs of externals with the smallest possible dimensions. Noting that the $AA \times TT $ OPE exchanges $AT$ and $TA$ at spin zero, and that $AA$ and $TT$ start in quadratic order in the fields perturbatively, the ideal system for studying these theories appears to be a $\phi$-$AA$-$TT$ mixed correlator system.

\subsection{Numerical results for the cubic theory}
Before proceeding to specific results, let us mention some details concerning our specific numerical setup. For the generation of crossing equations with operators transforming in a given global symmetry we use $\texttt{Autoboot}$ \cite{Go:2019lke,Go:2020ahx}. These are then imported into and optimised in $\texttt{Simpleboot}$ \cite{Simpleboot}, which sets up the numerical problem and communicates with $\texttt{SDPB}$ \cite{Simmons-Duffin:2015qma,Landry:2019qug} and other custom executables to provide the final bootstrap results we present in the paper. For lectures and many explicit examples for $\texttt{Simpleboot}$, we strongly recommend the lecture series \cite{PerimeterCourse}, along with the example notebooks shared at \cite{PerimeterCourseTutorials}.

In order to avoid clutter in the main text, we provide the assumptions and numerical parameters for each plot in the corresponding caption. See also Appendix \ref{AppendixParameters}. Another important comment regarding the gap assumptions for all plots in this work, is that they are comfortably in agreement with the multi-loop predictions for the spectrum of the $C_3$ theory provided in \cite{Bednyakov:2023lfj,Henriksson:2025hwi,Henriksson:2025vyi}. For every gap we have imposed in the present paper, we have cross-checked that the first operator after it in the cubic theory is comfortably allowed. For example, in some plots we will impose the assumption $\Delta_{T_{\mu\nu}^\prime}\geq 4$ on the first operator after the stress tensor, whereas the actual operator itself is found at $\Delta_{T_{\mu\nu}^\prime}=4.73644$.\footnote{This corresponds to a Padé$_{3,2}$ resummation. No error bar is implied by the number of significant digits retained.}

\subsubsection{\texorpdfstring{$X$}{X}-\texorpdfstring{$Z$}{Z} correlator system: the cubic redundancy channel}
We start by considering the simplest system of correlators that includes $X$ and $Z$ as externals. This is trivially the mixed $X$-$Z$ correlator system, which exchanges the desired $B$ irrep with a large gap at Lorentz spin zero. As we see in Figs.\ \ref{fig:ZXLambda11}, \ref{fig:ZXLambda19} and \ref{fig:ZXLambda27}, this system presents a feature that we refer to as the cubic redundancy channel. More specifically, we observe a strip of disallowed parameter space created precisely along the $\Delta_X=\Delta_Z$ symmetry enhancement line.\footnote{Both $X$ and $Z$ operators are of quadratic order in the fundamental field $\phi$ and stem from the branching of the $t \sim \phi_i \phi_j - \text{trace}$ operator of $O(3)$, which transforms in the rank-two symmetric traceless irrep. Hence, $\Delta_X=\Delta_Z$ represents a line in parameter space where the symmetry is enhanced from $C_3$ to $O(3)$.} Green points are allowed, red ones are not. The three figures correspond to increasing values of $\Lambda$ used, at the highest of which, $\Lambda=27$, the value of $\Delta_t=1.20954(32)$ \cite{Chester:2020iyt} of $O(3)$ has not yet been excluded. However, we observe that as we increase $\Lambda$, i.e.\ as we increase the constraining power of the algorithm, the channel is dug further and further into larger values of $\Delta_t$. 

We expect that for sufficiently large $\Lambda$, the values of $\Delta_t$ corresponding to the $O(3)$ fixed point will also be excluded. Indeed, we will soon see that including the operator $\phi$ in the system of external operators considered in our crossing equations, the channel dug out will reach the $O(3)$ theory. The intuition behind this is rather simple. The dimensions $\Delta_X$ and $\Delta_Z$ are rather large compared to $\Delta_\phi$, and so the system of correlators involving only $X$ and $Z$ is much less constraining than the one including also $\phi$. The numerical bootstrap is known to become progressively weaker as the dimensions of external operators becomes larger.\footnote{We note however that there has been been progress in solving this issue by using analytic functionals, which are distinct from the usual derivative functionals; see for example \cite{Paulos:2019gtx, Ghosh:2023onl}.}

\begin{figure}[H]
    \centering
    \begin{tikzpicture}
        \begin{axis}[
        xmin=0.9, xmax=1,
        ymin=0.9, ymax=1,
        xlabel= $\Delta_Z$,
        ylabel= $\Delta_X$,
        ylabel style={rotate=-90},
        xticklabel style={/pgf/number format/fixed, /pgf/number format/precision=3},
        xtick distance=0.01,
        ytick distance=0.01,
        minor x tick num=1,
        minor y tick num=1,
        xticklabel shift=0.1cm,
        yticklabel style={scaled ticks=false, /pgf/number format/fixed, /pgf/number format/precision=3}
        ]
        \addplot+[
            only marks,
            mark=*,
            mark options={color=Dark2-A,fill=Dark2-A},
            mark size=1pt] table{PlotData/ZX_Lambda11_allowed.dat};
        \addplot+[
            only marks,
            mark=*,
            mark options={color=Dark2-B,fill=Dark2-B},
            mark size=1pt] table{PlotData/ZX_Lambda11_not_allowed.dat};
        \addplot+[mark=none]
            coordinates {(0.9,0.9) (1,1)};
        \end{axis}
    \end{tikzpicture}
    \caption{Channel in the $\Delta_Z$-$\Delta_X$ plane obstructing symmetry enhancement, calculated at $\Lambda=11$. The numerical parameters used are \texttt{Set A} and \texttt{Set 1} of Appendix \ref{AppendixParameters}. The gap assumptions on the spectrum are $\Delta_S \geq 1.5$, $\Delta_{X^\prime}\geq2.8$, $\Delta_{Z^\prime}\geq2.8$, $\Delta_{\overline{X\lnsp X}_{\mu}}\geq 3.0$ and $\Delta_{T_{\mu\nu}^\prime} \geq 3.5$. Names and constructions of representations are given in \cite{Bednyakov:2023lfj}. Primes denote a subleading operator in a given sector. A twist gap of $\delta =10^{-6}$ is imposed on all operators not mentioned. We also impose that the ratio of OPE coefficients $\frac{\lambda_{XXT_{\mu \nu}}}{\lambda_{ZZT_{\mu\nu}}}=\frac{\Delta_X}{\Delta_Z}$ is fixed. All assumptions are comfortably in agreement with the calculations of \cite{Bednyakov:2023lfj,Henriksson:2025hwi,Henriksson:2025vyi}.}\label{fig:ZXLambda11}
\end{figure}
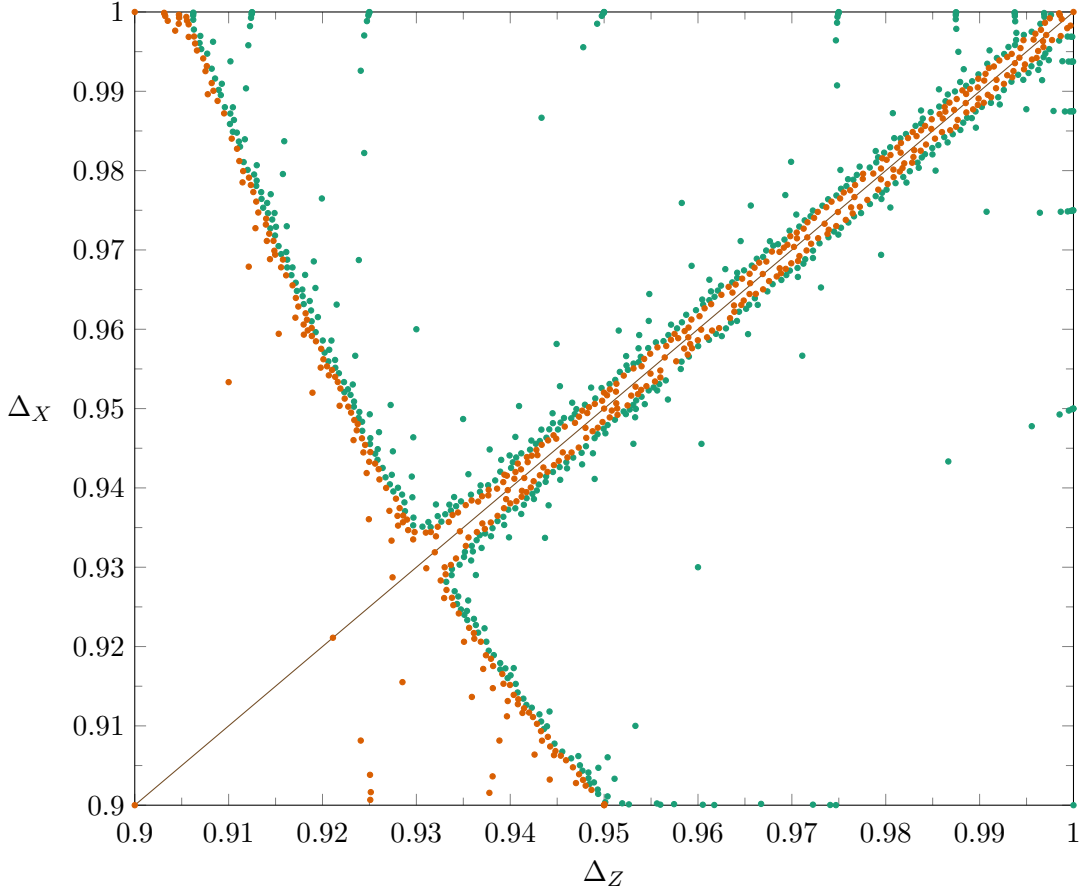

\begin{figure}[H]
    \centering
    \begin{tikzpicture}
        \begin{axis}[
        xmin=0.94, xmax=1.12,
        ymin=0.94, ymax=1.12,
        xlabel= $\Delta_Z$,
        ylabel= $\Delta_X$,
        ylabel style={rotate=-90},
        xticklabel style={/pgf/number format/fixed, /pgf/number format/precision=3},
        xtick distance=0.02,
        ytick distance=0.02,
        minor x tick num=1,
        minor y tick num=1,
        xticklabel shift=0.1cm,
        yticklabel style={scaled ticks=false, /pgf/number format/fixed, /pgf/number format/precision=3}
        ]
        \addplot+[
            only marks,
            mark=*,
            mark options={color=Dark2-A,fill=Dark2-A},
            mark size=1pt] table{PlotData/ZX_Lambda19_allowed.dat};
        \addplot+[
            only marks,
            mark=*,
            mark options={color=Dark2-B,fill=Dark2-B},
            mark size=1pt] table{PlotData/ZX_Lambda19_not_allowed.dat};
        \addplot+[mark=none]
            coordinates {(0.94,0.94) (1.12,1.12)};
        \end{axis}
    \end{tikzpicture}
    \caption{Channel in the $\Delta_Z$-$\Delta_X$ plane obstructing symmetry enhancement, calculated at $\Lambda=19$. The numerical parameters used are \texttt{Set B} and \texttt{Set 1} of Appendix \ref{AppendixParameters}. The gaps on the spectrum and other assumptions are as in Fig.\ \ref{fig:ZXLambda11}.}\label{fig:ZXLambda19}
\end{figure}
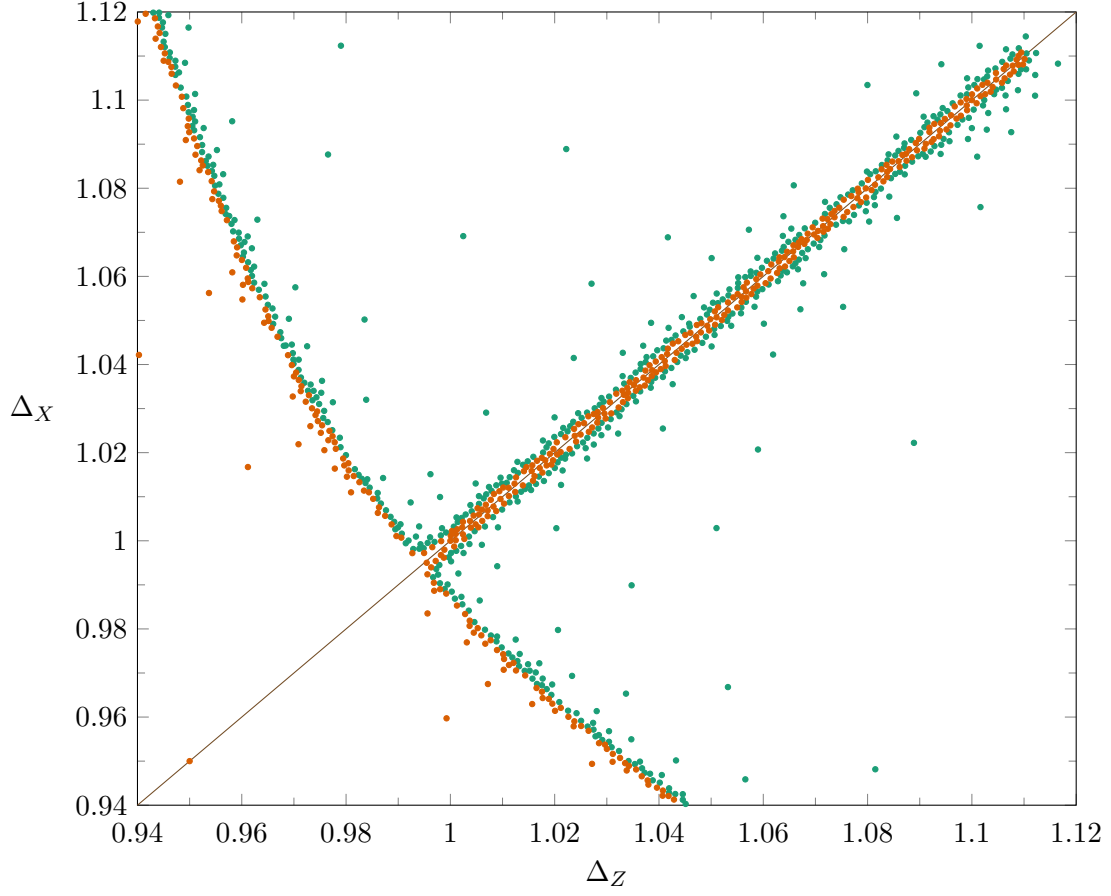

\begin{figure}[H]
    \centering
    \begin{tikzpicture}
        \begin{axis}[
        xmin=0.94, xmax=1.14,
        ymin=0.94, ymax=1.14,
        xlabel= $\Delta_Z$,
        ylabel= $\Delta_X$,
        ylabel style={rotate=-90},
        xticklabel style={/pgf/number format/fixed, /pgf/number format/precision=3},
        xtick distance=0.02,
        ytick distance=0.02,
        minor x tick num=1,
        minor y tick num=1,
        xticklabel shift=0.1cm,
        yticklabel style={scaled ticks=false, /pgf/number format/fixed, /pgf/number format/precision=3}
        ]
        \addplot+[
            only marks,
            mark=*,
            mark options={color=Dark2-A,fill=Dark2-A},
            mark size=1pt] table{PlotData/ZX_Lambda27_allowed.dat};
        \addplot+[
            only marks,
            mark=*,
            mark options={color=Dark2-B,fill=Dark2-B},
            mark size=1pt] table{PlotData/ZX_Lambda27_not_allowed.dat};
        \addplot+[mark=none]
            coordinates {(0.94,0.94) (1.14,1.14)};
        \end{axis}
    \end{tikzpicture}
    \caption{Channel in the $\Delta_Z$-$\Delta_X$ plane obstructing symmetry enhancement, calculated at $\Lambda=27$. The numerical parameters used are \texttt{Set B} and \texttt{Set 1} of Appendix \ref{AppendixParameters}. The gaps on the spectrum and other assumptions are as in Fig.\ \ref{fig:ZXLambda11}.}\label{fig:ZXLambda27}
\end{figure}
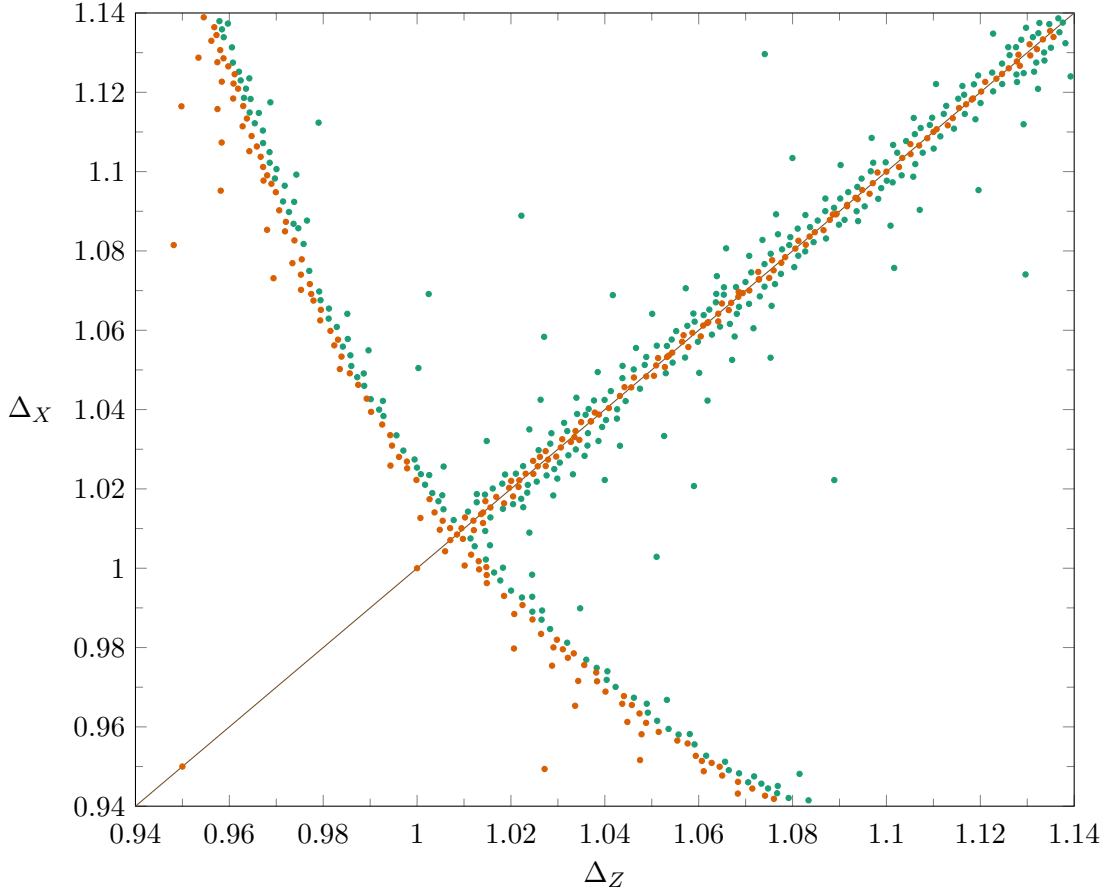
    
\subsubsection{\texorpdfstring{$\phi$-$Z$}{phi-Z} correlator system: obtaining an island}
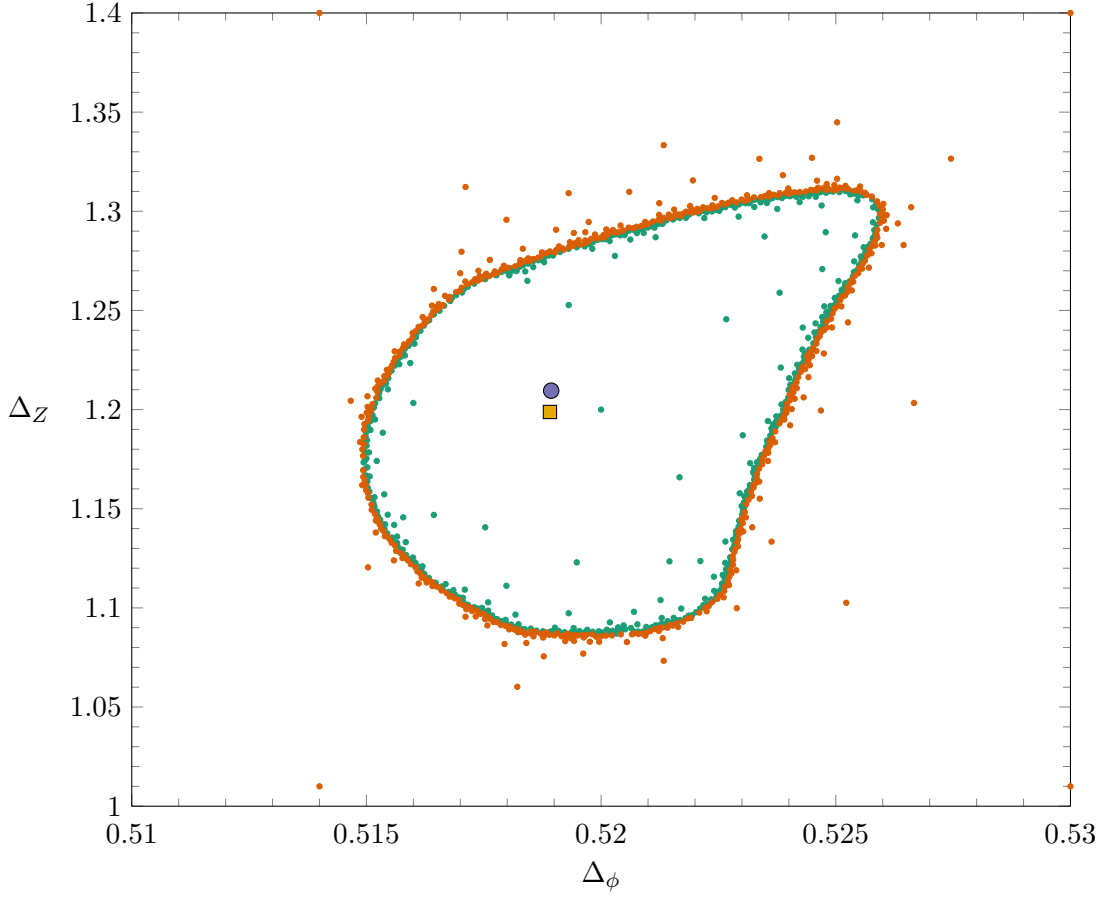
\begin{figure}[H]
    \centering
    \begin{tikzpicture}
        \begin{axis}[
        xmin=0.51, xmax=0.53,
        ymin=1, ymax=1.4,
        xlabel= $\Delta_\phi$,
        ylabel= $\Delta_Z$,
        ylabel style={rotate=-90},
        xticklabel style={/pgf/number format/fixed, /pgf/number format/precision=3},
        xtick distance=0.005,
        ytick distance=0.05,
        minor x tick num=4,
        minor y tick num=4,
        xticklabel shift=0.1cm,
        yticklabel style={scaled ticks=false, /pgf/number format/fixed, /pgf/number format/precision=2}
        ]
        \addplot+[
            only marks,
            mark=*,
            mark options={color=Dark2-A,fill=Dark2-A},
            mark size=1pt] table{PlotData/phiZ_Lambda19_allowed.dat};
        \addplot+[
            only marks,
            mark=*,
            mark options={color=Dark2-B,fill=Dark2-B},
            mark size=1pt] table{PlotData/phiZ_Lambda19_not_allowed.dat};
        \addplot+[
            mark=*, 
            mark options={scale=1.45,color=black,fill=Dark2-C}]
            coordinates {(0.518936,1.20954)};
        \addplot+[
            mark=square*, 
            mark options={scale=1.25,color=black,fill=Dark2-F}]
            coordinates {(0.51891,1.1988)};
        \end{axis}
    \end{tikzpicture}
    \caption{Island in the $\Delta_\phi$-$\Delta_Z$ plane at $\Lambda = 19$. The island contains both the $C_3$ and $O(3)$ CFTs. This will be remedied when we study the full $\phi$-$X$-$Z$ correlator system. The numerical parameters used are \texttt{Set B} and \texttt{Set 1} of Appendix \ref{AppendixParameters}. The gap assumptions on the spectrum are $\Delta_S \geq 1.5$, $\Delta_{Z^\prime}\geq2.8$, $\Delta_{\phi^\prime}\geq 1.5$, $\Delta_{Z_3}\geq 1.5$, $\Delta_{XV}\geq 1.5$ and $\Delta_{T_{\mu\nu}^\prime} \geq 4.0$. Names and constructions of representations are given in \cite{Bednyakov:2023lfj}.  Primes denote a subleading operator in a given sector. A twist gap of $\delta =10^{-6}$ is imposed on all operators not mentioned, except for the leading $B$ spin-one operator (the would be $O(3)$ current). We also impose that the ratio of OPE coefficients $\frac{\lambda_{\phi \phi T_{\mu \nu}}}{\lambda_{ZZT_{\mu\nu}}}=\frac{\Delta_\phi}{\Delta_Z}$ is fixed. All assumptions are comfortably in agreement with the calculations of \cite{Bednyakov:2023lfj,Henriksson:2025hwi,Henriksson:2025vyi}. The blue circle represents the central value of the $O(3)$ bootstrap determination \cite{Chester:2020iyt}, $(\Delta_\phi, \Delta_t) =(0.518936(67), 1.20954(32))$. The $O(3)$ theory lies deep in the allowed region, and without our analysis on redundant operators would be particularly hard to exclude. The yellow square gives the location of the $C_3$ theory using the central values of the results $\Delta_\phi=0.51891(7)$ from \cite{Hasenbusch:2022zur} and $\Delta_Z=1.1988(24)$ from \cite{Rong:2023owx}.}\label{fig:PhiZLambda19}
\end{figure}

\begin{figure}[H]
    \centering
    \begin{tikzpicture}
        \begin{axis}[
        xmin=0.51, xmax=0.53,
        ymin=1, ymax=1.4,
        xlabel= $\Delta_\phi$,
        ylabel= $\Delta_Z$,
        ylabel style={rotate=-90},
        xticklabel style={/pgf/number format/fixed, /pgf/number format/precision=3},
        xtick distance=0.005,
        ytick distance=0.05,
        minor x tick num=4,
        minor y tick num=4,
        xticklabel shift=0.1cm,
        yticklabel style={scaled ticks=false, /pgf/number format/fixed, /pgf/number format/precision=2}
        ]
        \addplot+[
            only marks,
            mark=*,
            mark options={color=Dark2-A,fill=Dark2-A},
            mark size=1pt] table{PlotData/phiZ_stronger_Lambda19_allowed.dat};
        \addplot+[
            only marks,
            mark=*,
            mark options={color=Dark2-B,fill=Dark2-B},
            mark size=1pt] table{PlotData/phiZ_stronger_Lambda19_not_allowed.dat};
        \addplot+[
            mark=*, 
            mark options={scale=1.45,color=black,fill=Dark2-C}]
            coordinates {(0.518936,1.20954)};
        \addplot+[
            mark=square*, 
            mark options={scale=1.25,color=black,fill=Dark2-F}]
            coordinates {(0.51891,1.1988)};
        \end{axis}
    \end{tikzpicture}
    \caption{Island in the $\Delta_\phi$-$\Delta_Z$ plane at $\Lambda = 19$, with stronger assumptions compared to Fig.\ \ref{fig:PhiZLambda19}. The island contains both the $C_3$ and $O(3)$ CFTs. This will be remedied when we study the full $\phi$-$X$-$Z$ correlator system. The numerical parameters used are \texttt{Set B} and \texttt{Set 1} of Appendix \ref{AppendixParameters}. The gap assumptions on the spectrum are $\Delta_S \geq 1.5$, $\Delta_X \geq 1.1$, $\Delta_{Z^\prime}\geq2.8$, $\Delta_{\phi^\prime}\geq 1.8$, $\Delta_{Z_3}\geq 1.8$, $\Delta_{XV}\geq 1.8$, $\Delta_{VB} \geq 3.0$ and $\Delta_{T_{\mu\nu}^\prime} \geq 4.0$. Names and constructions of representations are given in \cite{Bednyakov:2023lfj}. Primes denote a subleading operator in a given sector. A twist gap of $\delta =10^{-6}$ is imposed on all operators not mentioned, except for the leading $B$ spin-one operator (the would be $O(3)$ current). We also impose that the ratio of OPE coefficients $\frac{\lambda_{\phi \phi T_{\mu \nu}}}{\lambda_{ZZT_{\mu\nu}}}=\frac{\Delta_\phi}{\Delta_Z}$ is fixed. All assumptions are comfortably in agreement with the calculations of \cite{Bednyakov:2023lfj,Henriksson:2025hwi,Henriksson:2025vyi}. The blue circle represents the central value of the $O(3)$ bootstrap determination \cite{Chester:2020iyt}, $(\Delta_\phi, \Delta_t) =(0.518936(67), 1.20954(32))$. The $O(3)$ theory lies deep in the allowed region, and without our analysis on redundant operators would be particularly hard to exclude. The yellow square gives the location of the $C_3$ theory using the central values of the results $\Delta_\phi=0.51891(7)$ from \cite{Hasenbusch:2022zur} and $\Delta_Z=1.1988(24)$ from \cite{Rong:2023owx}.}\label{fig:PhiZLambda19MoreGaps}
\end{figure}
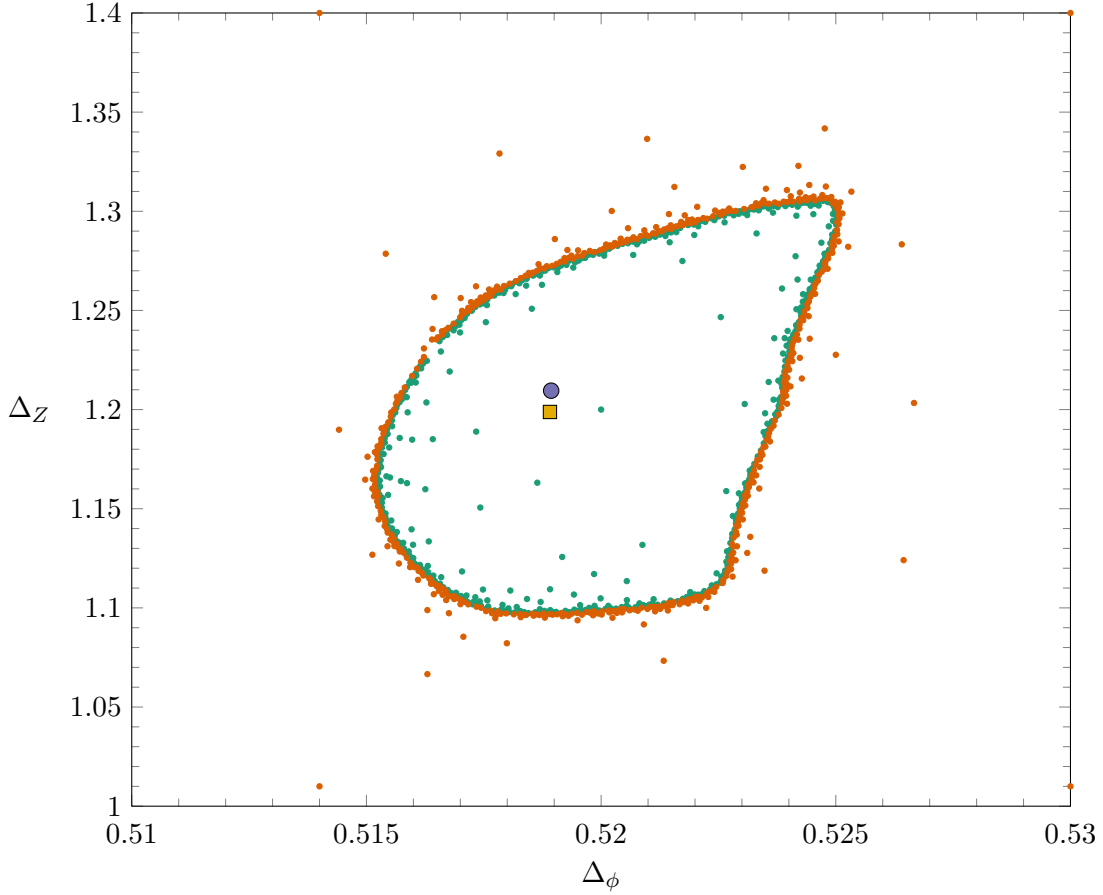

Having observed that the $X$-$Z$ mixed correlator system is capable of excluding the $O(3)$ symmetry enhanced line from parameter space, we would now like to find a system capable of also providing an island. This is necessary if we want to be able to perform a precision study of the theory. We find that the $\phi$-$Z$ correlator system is indeed capable of doing this. In Figs.\ \ref{fig:PhiZLambda19} and \ref{fig:PhiZLambda19MoreGaps}, which correspond to two sets of gaps imposed on the spectrum (see the corresponding captions), we find our desired island. The main assumption necessary for obtaining the island is the gap on the subleading operator in the singlet Lorentz spin two channel, after the stress-energy tensor itself. In particular, we impose $\Delta_{T_{\mu\nu}^\prime}\geq 4$.

\subsubsection{\texorpdfstring{$\phi$-$X$-$Z$}{phi-X-Z} correlator system: obtaining an island without symmetry enhancement}
\begin{figure}[H]
    \centering
    \begin{tikzpicture}
        \begin{axis}[
        xmin=1, xmax=1.4,
        ymin=1, ymax=1.4,
        xlabel= $\Delta_X$,
        ylabel= $\Delta_Z$,
        ylabel style={rotate=-90},
        xticklabel style={/pgf/number format/fixed, /pgf/number format/precision=3},
        xtick distance=0.05,
        ytick distance=0.05,
        minor x tick num=4,
        minor y tick num=4,
        xticklabel shift=0.1cm,
        yticklabel style={scaled ticks=false, /pgf/number format/fixed, /pgf/number format/precision=2}
        ]
        \addplot+[
            only marks,
            mark=*,
            mark options={color=Dark2-A,fill=Dark2-A},
            mark size=1pt] table{PlotData/XZ_Lambda19_0518936_allowed.dat};
        \addplot+[
            only marks,
            mark=*,
            mark options={color=Dark2-B,fill=Dark2-B},
            mark size=1pt] table{PlotData/XZ_Lambda19_0518936_not_allowed.dat};
        \addplot+[mark=none]
            coordinates {(1,1) (1.4,1.4)};
        \addplot+[
            mark=*, 
            mark options={scale=1.65,color=black,fill=Dark2-C}]
            coordinates {(1.20954,1.20954)};
        \addplot+[
            mark=square*, 
            mark options={scale=1.5,color=black,fill=Dark2-F}]
            coordinates {(1.2256,1.1988)};  
        \end{axis}
    \end{tikzpicture}
    \caption{Projection onto the $\Delta_X$-$\Delta_Z$ plane at $\Delta_\phi =0.51893$ of the full $\Delta_\phi$-$\Delta_X$-$\Delta_Z$ island, at $\Lambda = 19$. The value of $\Delta_\phi$ is taken to be the central value reported in \cite{Chester:2020iyt} for the $O(3)$ model. The numerical parameters used are \texttt{Set B} and \texttt{Set 2} of Appendix \ref{AppendixParameters}.  The gap assumptions on the spectrum are $\Delta_S \geq 1.5$, $\Delta_{X^\prime} \geq 2.5$, $\Delta_{Z^\prime}\geq2.5$, $\Delta_{\phi^\prime}\geq 1.5$, $\Delta_{B}\geq 4.0$ and $\Delta_{T_{\mu\nu}^\prime} \geq 4.0$. Names and constructions of representations are given in \cite{Bednyakov:2023lfj}. Primes denote a subleading operator in a given sector. A twist gap of $\delta =10^{-6}$ is imposed on all operators not mentioned. We also impose that the ratios of OPE coefficients $\frac{\lambda_{\phi \phi T_{\mu \nu}}}{\lambda_{ZZT_{\mu\nu}}}=\frac{\Delta_\phi}{\Delta_Z}$ and $\frac{\lambda_{X X T_{\mu \nu}}}{\lambda_{ZZT_{\mu\nu}}}=\frac{\Delta_X}{\Delta_Z}$ are fixed. All assumptions are comfortably in agreement with the calculations of \cite{Bednyakov:2023lfj,Henriksson:2025hwi,Henriksson:2025vyi}. The yellow square represents the central value of the conformal perturbation theory results from \cite{Rong:2023owx}, $(\Delta_X, \Delta_Z) =(1.2256(36),1.1988(24))$. The blue circle is the central value of the $O(3)$ theory as given in \cite{Chester:2020iyt}. A channel along the diagonal due to the assumption $\Delta_B \geq 4.0$ is observed.}\label{fig:PhiXZLambda19Channel}
\end{figure}
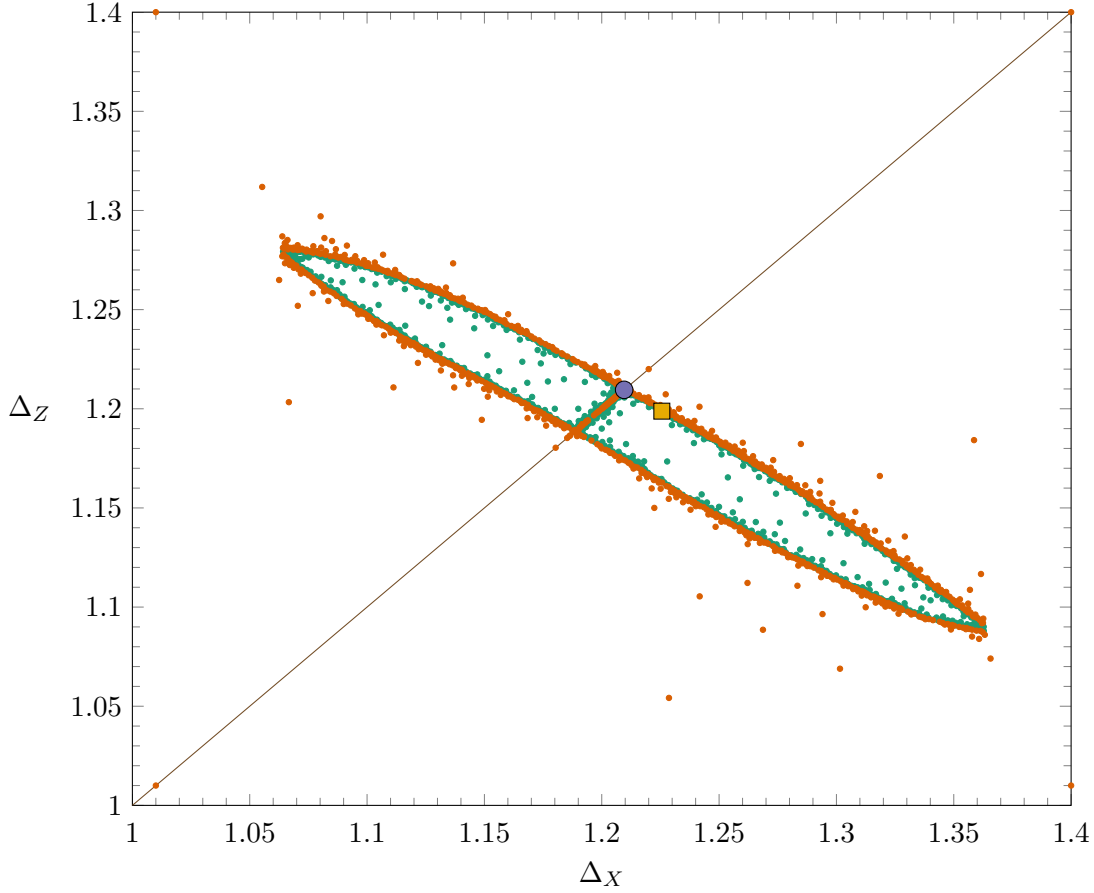

We have thus far learnt that the $Z$-$X$ system is capable of preventing symmetry enhancement, and that the $\phi$-$Z$ system is capable of producing and island. Thus, studying a $\phi$-$X$-$Z$ mixed correlator system should be sufficient to produce an island that does not include in it the $O(3)$ fixed point. In Fig.\ \ref{fig:PhiXZLambda19Channel}, we see that this is indeed the case. This figure corresponds to a two-dimensional slice of a larger three-dimensional island we have obtained. This slice is taken precisely at the value of $\Delta_\phi$ that corresponds to the $O(3)$ fixed point, i.e.\ $\Delta_\phi=0.51893$ \cite{Chester:2020iyt}.

\begin{figure}[H]
    \centering
    \begin{tikzpicture}
        \begin{axis}[
        xmin=1, xmax=1.4,
        ymin=1, ymax=1.4,
        xlabel= $\Delta_X$,
        ylabel= $\Delta_Z$,
        ylabel style={rotate=-90},
        xticklabel style={/pgf/number format/fixed, /pgf/number format/precision=3},
        xtick distance=0.05,
        ytick distance=0.05,
        minor x tick num=4,
        minor y tick num=4,
        xticklabel shift=0.1cm,
        yticklabel style={scaled ticks=false, /pgf/number format/fixed, /pgf/number format/precision=2}
        ]
        \addplot+[
            only marks,
            mark=*,
            mark options={color=Dark2-A,fill=Dark2-A},
            mark size=1pt] table{PlotData/XZ_Lambda19_05185_allowed.dat};
        \addplot+[
            only marks,
            mark=*,
            mark options={color=Dark2-B,fill=Dark2-B},
            mark size=1pt] table{PlotData/XZ_Lambda19_05185_not_allowed.dat};
        \addplot+[mark=none]
            coordinates {(1,1) (1.4,1.4)};
        \addplot+[
            mark=square*, 
            mark options={scale=1.5,color=black,fill=Dark2-F}]
            coordinates {(1.2256,1.1988)};       
        \end{axis}
    \end{tikzpicture}
    \caption{Projection onto the $\Delta_X$-$\Delta_Z$ plane at $\Delta_\phi=0.5185$ of the full $\Delta_\phi$-$\Delta_X$-$\Delta_Z$ island, at $\Lambda=19$. The numerical parameters used are \texttt{Set B} and \texttt{Set 2} of Appendix \ref{AppendixParameters}.  The gap assumptions on the spectrum are $\Delta_S \geq 1.5$, $\Delta_{X^\prime} \geq 2.5$, $\Delta_{Z^\prime}\geq2.5$, $\Delta_{\phi^\prime}\geq 1.5$, $\Delta_{B}\geq 4.0$ and $\Delta_{T_{\mu\nu}^\prime} \geq 4.0$. Names and constructions of representations are given in \cite{Bednyakov:2023lfj}. Primes denote a subleading operator in a given sector. A twist gap of $\delta =10^{-10}$ is imposed on all operators not mentioned. We also impose that the ratios of OPE coefficients $\frac{\lambda_{\phi \phi T_{\mu \nu}}}{\lambda_{ZZT_{\mu\nu}}}=\frac{\Delta_\phi}{\Delta_Z}$ and $\frac{\lambda_{X X T_{\mu \nu}}}{\lambda_{ZZT_{\mu\nu}}}=\frac{\Delta_X}{\Delta_Z}$ are fixed. All assumptions are comfortably in agreement with the calculations of \cite{Bednyakov:2023lfj,Henriksson:2025hwi,Henriksson:2025vyi}. The yellow square represents the central value of the conformal perturbation theory results from \cite{Rong:2023owx}, $(\Delta_X, \Delta_Z) =(1.2256(36),1.1988(24))$. A channel along the diagonal due to the assumption $\Delta_B \geq 4.0$ is observed.}\label{fig:PhiXZLambda19Channel05185}
\end{figure}
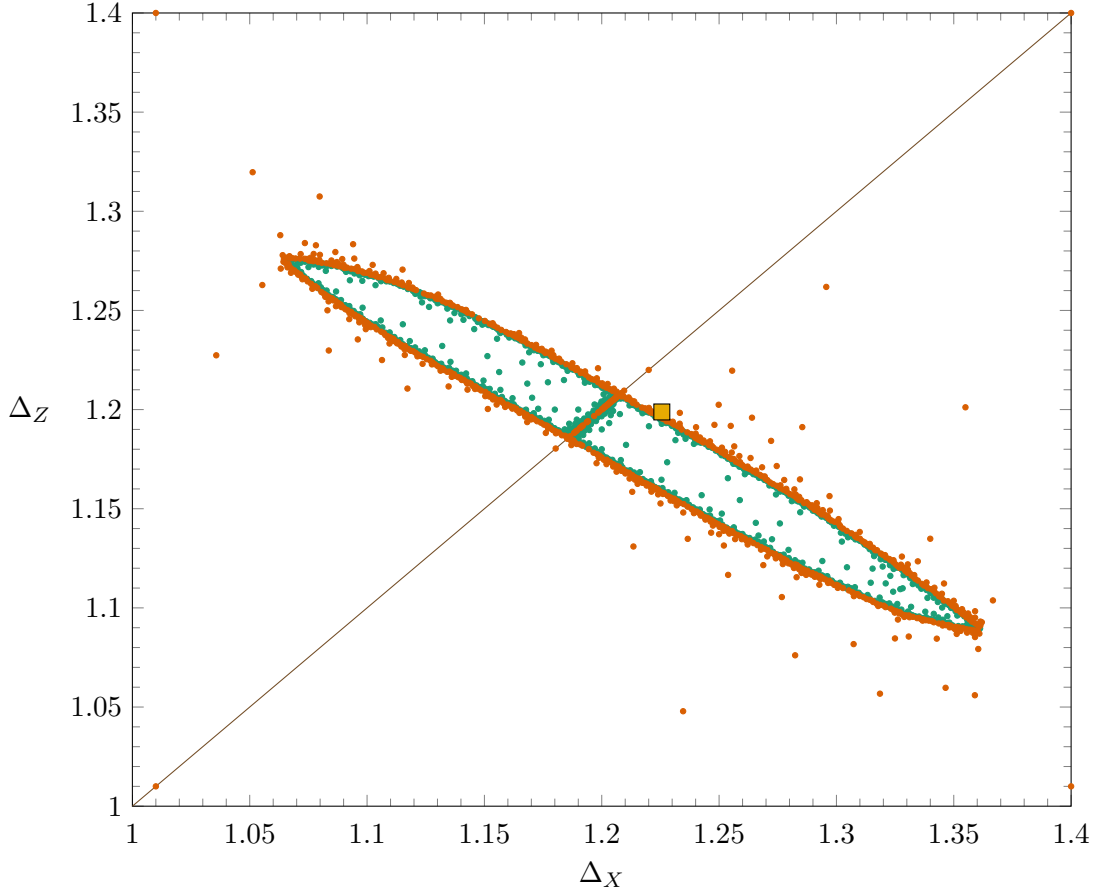

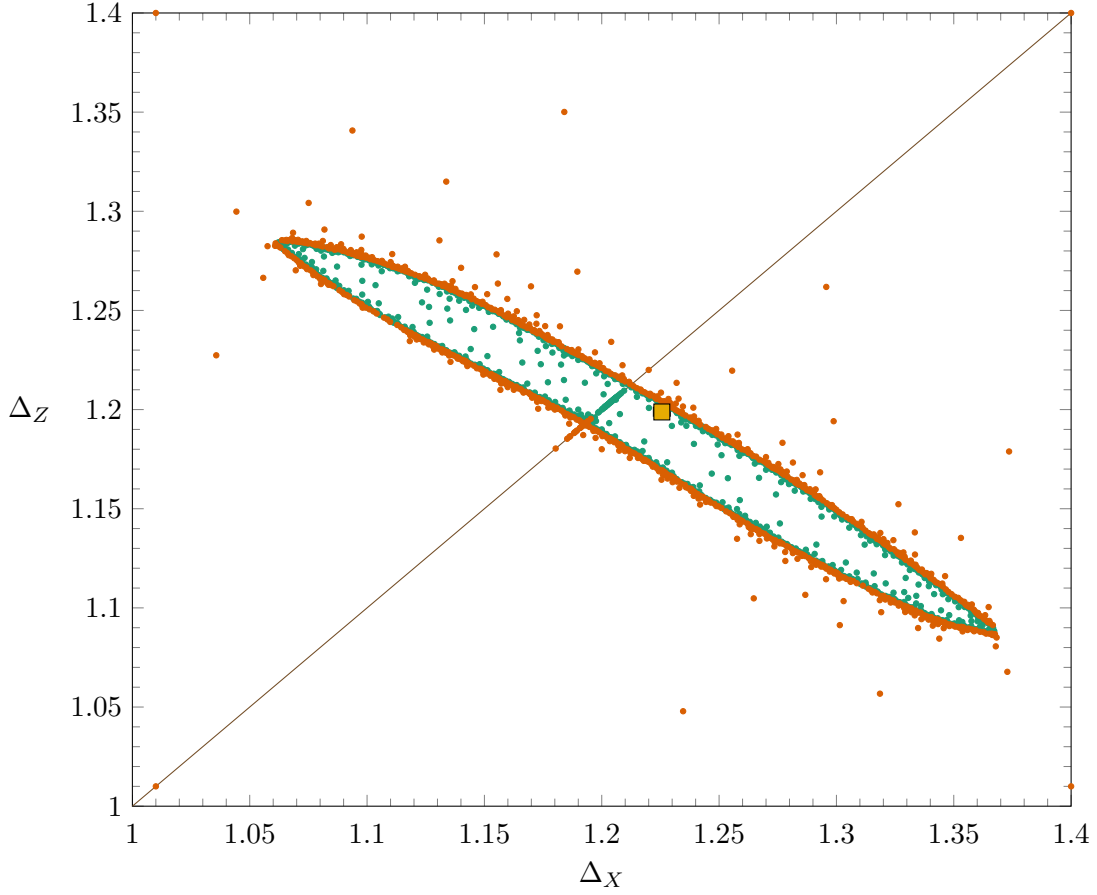
\begin{figure}[H]
    \centering
    \begin{tikzpicture}
        \begin{axis}[
        xmin=1, xmax=1.4,
        ymin=1, ymax=1.4,
        xlabel= $\Delta_X$,
        ylabel= $\Delta_Z$,
        ylabel style={rotate=-90},
        xticklabel style={/pgf/number format/fixed, /pgf/number format/precision=3},
        xtick distance=0.05,
        ytick distance=0.05,
        minor x tick num=4,
        minor y tick num=4,
        xticklabel shift=0.1cm,
        yticklabel style={scaled ticks=false, /pgf/number format/fixed, /pgf/number format/precision=2}
        ]
        \addplot+[
            only marks,
            mark=*,
            mark options={color=Dark2-A,fill=Dark2-A},
            mark size=1pt] table{PlotData/XZ_Lambda19_05195_allowed.dat};
        \addplot+[
            only marks,
            mark=*,
            mark options={color=Dark2-B,fill=Dark2-B},
            mark size=1pt] table{PlotData/XZ_Lambda19_05195_not_allowed.dat};
        \addplot+[mark=none]
            coordinates {(1,1) (1.4,1.4)};
        \addplot+[mark=none]
            coordinates {(1.222,1.1964) (1.222,1.2012) (1.2292,1.2012) (1.2292,1.1964)};
        \addplot+[
            mark=square*, 
            mark options={scale=1.5,color=black,fill=Dark2-F}]
            coordinates {(1.2256,1.1988)};  
        \end{axis}
    \end{tikzpicture}
    \caption{Projection onto the $\Delta_X$-$\Delta_Z$ plane at $\Delta_\phi=0.5195$ of the full $\Delta_\phi$-$\Delta_X$-$\Delta_Z$ island, at $\Lambda=19$. The numerical parameters used are \texttt{Set B} and \texttt{Set 2} of Appendix \ref{AppendixParameters}.  The gap assumptions on the spectrum are $\Delta_S \geq 1.5$, $\Delta_{X^\prime} \geq 2.5$, $\Delta_{Z^\prime}\geq2.5$, $\Delta_{\phi^\prime}\geq 1.5$, $\Delta_{B}\geq 4.0$ and $\Delta_{T_{\mu\nu}^\prime} \geq 4.0$. Names and constructions of representations are given in \cite{Bednyakov:2023lfj}. Primes denote a subleading operator in a given sector. A twist gap of $\delta =10^{-10}$ is imposed on all operators not mentioned. We also impose that the ratios of OPE coefficients $\frac{\lambda_{\phi \phi T_{\mu \nu}}}{\lambda_{ZZT_{\mu\nu}}}=\frac{\Delta_\phi}{\Delta_Z}$ and $\frac{\lambda_{X X T_{\mu \nu}}}{\lambda_{ZZT_{\mu\nu}}}=\frac{\Delta_X}{\Delta_Z}$ are fixed. All assumptions are comfortably in agreement with the calculations of \cite{Bednyakov:2023lfj,Henriksson:2025hwi,Henriksson:2025vyi}. The yellow square represents the central value of the conformal perturbation theory results from \cite{Rong:2023owx}, $(\Delta_X, \Delta_Z) =(1.2256(36),1.1988(24))$. Here we observe that a channel along the diagonal due to $\Delta_B \geq 4.0$ is only partially opened at the bottom side of the island, and does not penetrate deep into it.}\label{fig:PhiXZLambda19Channel05195}
\end{figure}

In Fig.\ \ref{fig:PhiXZLambda19NoChannel}, we also observe that removing the assumption $\Delta_B \geq 4$, i.e.\ the assumption that prohibits symmetry enhancement, also removes the redundancy channel, as expected. In Figs.\ \ref{fig:PhiXZLambda19Channel05185} and \ref{fig:PhiXZLambda19Channel05195} we present two additional slices, at fixed $\Delta_\phi =0.5185$ and $\Delta_\phi=0.5195$, respectively. For $\Delta_\phi=0.5185$ we observe that the channel separates the island in two, as in the $\Delta_\phi=\Delta_\phi^{O(3)}$ case, whereas for $\Delta_\phi= 0.5195$ a channel has only partially started being dug out.

\begin{figure}[H]
    \centering
    \begin{tikzpicture}
        \begin{axis}[
        xmin=1, xmax=1.4,
        ymin=1, ymax=1.4,
        xlabel= $\Delta_X$,
        ylabel= $\Delta_Z$,
        ylabel style={rotate=-90},
        xticklabel style={/pgf/number format/fixed, /pgf/number format/precision=3},
        xtick distance=0.05,
        ytick distance=0.05,
        minor x tick num=4,
        minor y tick num=4,
        xticklabel shift=0.1cm,
        yticklabel style={scaled ticks=false, /pgf/number format/fixed, /pgf/number format/precision=2}
        ]
        \addplot+[
            only marks,
            mark=*,
            mark options={color=Dark2-A,fill=Dark2-A},
            mark size=1pt] table{PlotData/XZ_Lambda19_0518936_noB_allowed.dat};
        \addplot+[
            only marks,
            mark=*,
            mark options={color=Dark2-B,fill=Dark2-B},
            mark size=1pt] table{PlotData/XZ_Lambda19_0518936_noB_not_allowed.dat};
        \addplot+[mark=none]
            coordinates {(1,1) (1.4,1.4)};
        \addplot+[
            mark=*, 
            mark options={scale=1.65,color=black,fill=Dark2-C}]
            coordinates {(1.20954,1.20954)};
        \addplot+[
            mark=square*, 
            mark options={scale=1.5,color=black,fill=Dark2-F}]
            coordinates {(1.2256,1.1988)};         
        \end{axis}
    \end{tikzpicture}
    \caption{Projection onto the $\Delta_X$-$\Delta_Z$ plane at $\Delta_\phi=0.51893$ of the full $\Delta_\phi$-$\Delta_X$-$\Delta_Z$ island, at $\Lambda = 19$. The value of $\Delta_\phi$ is taken to be the central value reported in \cite{Chester:2020iyt} for the $O(3)$ model. The numerical parameters used are \texttt{Set B} and \texttt{Set 2} of Appendix \ref{AppendixParameters}. The gap assumptions on the spectrum are $\Delta_S \geq 1.5$, $\Delta_{X^\prime} \geq 2.5$, $\Delta_{Z^\prime}\geq2.5$, $\Delta_{\phi^\prime}\geq 1.5$ and $\Delta_{T_{\mu\nu}^\prime} \geq 4.0$. Names and constructions of representations are given in \cite{Bednyakov:2023lfj}. Primes denote a subleading operator in a given sector. A twist gap of $\delta =10^{-6}$ is imposed on all operators not mentioned. We also impose that the ratios of OPE coefficients $\frac{\lambda_{\phi \phi T_{\mu \nu}}}{\lambda_{ZZT_{\mu\nu}}}=\frac{\Delta_\phi}{\Delta_Z}$ and $\frac{\lambda_{X X T_{\mu \nu}}}{\lambda_{ZZT_{\mu\nu}}}=\frac{\Delta_X}{\Delta_Z}$ are fixed. All assumptions are comfortably in agreement with the calculations of \cite{Bednyakov:2023lfj,Henriksson:2025hwi,Henriksson:2025vyi}. The yellow square represents the central value of the conformal perturbation theory results from \cite{Rong:2023owx}, $(\Delta_X, \Delta_Z) =(1.2256(36),1.1988(24))$. The blue circle is the central value of the $O(3)$ theory as given in \cite{Chester:2020iyt}. Comparing with Fig.\ \ref{fig:PhiXZLambda19Channel}, we notice the absence of a channel along the diagonal due to the fact that we have not imposed $\Delta_B \geq 4.0$.}\label{fig:PhiXZLambda19NoChannel}
\end{figure}
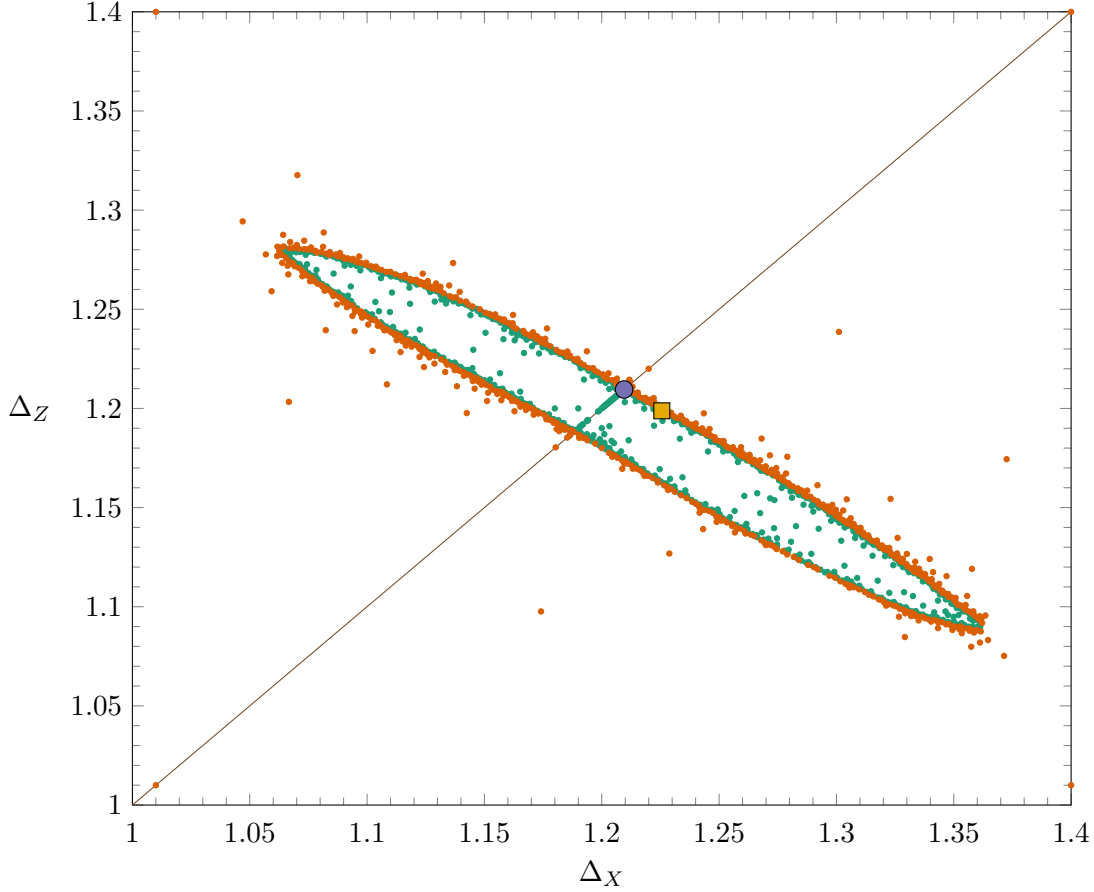

\section{Conclusion and future directions}\label{sec:conc}
In this work we have used inspiration from redundant operators of Lagrangian field theory to identify large gaps that can be used in the numerical conformal bootstrap. These gaps can be used as distinguishing characteristics for theories under study, allowing their isolation in the space of theories. A number of examples were provided, and we observed that different types of primary operators can disappear due to equation of motion effects leaving behind large gaps. Such operators include $\phi^3$ (Ising, $O(N)$), $\phi^4$ ($C_N$, $S_{N+1}\times \mathbb{Z}_2$, MN, $O(m)\times O(n)/\mathbb{Z}_2$) and $\partial \phi^4$ ($C_N$, MN) operators. The operators removed can also be seen as a consequence of non-conservation equations of the type $\partial_\mu J^{\mu \nu\ldots} - \mathcal{O}^{\nu\ldots}=0$. These redundant operators allow us to distinguish theories with different symmetries, as well as decoupled from coupled theories. Another example that we did not discuss in detail here is that of theories with $U(m)\times U(n)/U(1)$ global symmetry \cite{Kousvos:2022ewl}.

Our methods can also be applied to scalar-fermion theories, for which some islands have already been studied in \cite{Rong:2018okz, Atanasov:2018kqw, Atanasov:2022bpi, Erramilli:2022kgp, Mitchell:2024hix}. The present work provides the means to study theories of this type that are not $O(N)$ symmetric. The $\varepsilon$ expansion suggests the existence of a vast space of such fixed points \cite{Pannell:2023tzc}. Similarly, there is no obstruction to applying our methodology to gauge theories; however, in this case the main difficulty is expected to be the fact that even the lowest-lying operators have rather high dimension, making the numerics particularly demanding compared to pure scalar and scalar-fermion theories. The spectra of gauge theories within the context of the conformal bootstrap have been studied in \cite{He:2021xvg}.

Cases where spacetime symmetries are of interest, e.g.\ when a crossing equation obtained assuming some amount of supersymmetry ends up being satisfied by a theory with more supersymmetry, can also be similarly treated, using the breaking of the supercurrent and/or R-symmetry current. 

Another realm where our methods may be applied is in two-dimensional CFTs, when higher-spin conserved currents are broken. One example is the flow from $N$-decoupled three-state Potts models, with $S_3{\lnsp}^n\rtimes S_n$ global symmetry, to a fully coupled theory with the same global symmetry \cite{Dotsenko:1998gyp}. The decoupled theory possesses a spin-three conserved current, which in the coupled theory is believed to break (since the coupled theory is believed to be non-integrable).\footnote{Notably, from the point of view of global conformal symmetry (i.e.\ not Virasoro), there will also be a reduction in the number of stress-energy tensors from $n$ to $1$ when flowing from the decoupled models to the coupled theory. This is expected to be useful when the theory is studied using the global conformal bootstrap.} A bootstrap study of these theories was provided in \cite{Kousvos:2024dlz}. Similarly, the authors of \cite{Antunes:2022vtb, Antunes:2024mfb} have provided another class of examples, the simplest of which is a theory with $\mathbb{Z}_2{\!}^4\rtimes S_4$ global symmetry.\footnote{SRK thanks Antonio Antunes for pointing this out.} This can be thought of as four coupled replicas of the tricritical Ising model. The tricritical Ising model itself is obtainable in a $3-\varepsilon$ expansion ($\phi^6$ theory), and in \cite{Henriksson:2025kws} it was shown that an island in $d<3$ (but not yet in $d=2$) can be obtained. Studying such a $\mathbb{Z}_2{\!}^4\rtimes S_4$ symmetric theory in $d=2$ would thus be a worthwhile effort, aiming to shed further light on non-integrable 2D CFTs.

With respect to the numerical aspects of our work concerning the bootstrap of the $C_3$ theory in $d=3$, we have obtained an island for it that does not include the $O(3)$ theory. This resolves a problem that has been outstanding for a long time, and our solution paves the way for a precision numerical bootstrap study of this theory. We have already performed a preliminary study using the navigator method of \cite{Reehorst:2021ykw}, and scanning over the OPE coefficient ratios $(\frac{\lambda_{\phi \phi X}}{\lambda_{Z Z Z}},\frac{\lambda_{\phi \phi Z}}{\lambda_{Z Z Z}},\frac{\lambda_{X X X}}{\lambda_{Z Z Z}},\frac{\lambda_{Z Z X}}{\lambda_{Z Z Z}})$ did not provide any dramatic effect, at least at $\Lambda=17$ that we explicitly tested.\footnote{We also ran some tests using the skydiving method \cite{Liu:2023elz}, however we found it particularly unstable without fine tuning parameters.} We expect that once we raise the gaps imposed on the spectrum and/or mix in the leading scalar singlet ($S\sim\phi^2$) as an external operator, the OPE scan will start producing strong results, and the navigator method will prove necessary.\footnote{It is not a priori obvious to us what the optimal combination of methods will be to perform a precision study of the $C_3$ theory. The recent high-precision study \cite{Chang:2024whx} did not use the navigator method. Other studies, e.g.\ \cite{Mitchell:2024hix} and \cite{Henriksson:2025kws}, used the navigator but not the skydiving method.} Note also that mixing in $\phi^2$ as an external operator is expected to make a gap to the dimension of the first $\phi^4$ singlet operator have a substantial effect in the numerics. This operator is expected to satisfy a rigorous bound $\Delta>3$, but its dimension is expected to be extremely close to $\Delta=3$, as was seen in a recent Monte Carlo study \cite{Hasenbusch:2022zur}. This means that imposing a gap of $\Delta \geq 3$ should prove to be quite constraining, and that precision exponents for the $C_3$ theory can be obtained with the numerical conformal bootstrap.

\ack{We would like to acknowledge helpful conversations with Johan Henriksson, Petr Kravchuk, Marco Meineri, Alessandro Piazza, Jasper Roosmale Nepveu, Ning Su and Alessandro Vichi. SRK would like to acknowledge his co-authors on \cite{Bednyakov:2023lfj,Henriksson:2025hwi,Henriksson:2025vyi} for collaboration and discussions on work that proved central for the present paper. SRK acknowledges AS for hospitality and support during a visit to King's College London. 

SRK has received funding from the European Research Council (ERC) under the European Union's Horizon 2020 research and innovation programme (grant agreement no.\ 758903), and the Marie Skłodowska-Curie Action (MSCA) High energy Intelligence (HORIZON-MSCA-2023-SE-01-101182937-HeI). This research was supported in part by Perimeter Institute for Theoretical Physics. Research at Perimeter Institute is supported by the Government of Canada through the Department of Innovation, Science and Economic Development and by the Province of Ontario through the Ministry of Research and Innovation. AS is supported by the Royal Society under grant URF\textbackslash{}R1\textbackslash211417 and by STFC under grant ST/X000753/1.

The numerical computations in this work have used King's College London's CREATE~\cite{CREATE} computing cluster.}

\begin{appendices}

\section{Numerical parameters for plots}\label{AppendixParameters}
Here we briefly summarise the set of parameters we use in \texttt{Simpleboot} and \texttt{SDPB}. With regards to \texttt{Simpleboot}, we use three sets of parameters. These are \texttt{Set A}: $d=3$, $\Lambda=11$, $\kappa=12$, $r_N=48$ and  $\ell_{\text{set}} = \{0,\ldots,27, 49, 50\}$, \texttt{Set B}: $d=3$, $\Lambda=19$, $\kappa=14$, $r_N=56$ and  $\ell_{\text{set}} = \{0,\ldots,26, 49, 50\}$ and \texttt{Set C}:  $d=3$, $\Lambda=27$, $\kappa=20$, $r_N=80$ and  $\ell_{\text{set}} = \{0,\ldots,26,29,30,33,34,37,38,41,42,45,46,49,50\}$. In brief, $\Lambda$ controls the number of functionals, i.e.\ the number of derivatives we take of the crossing equation, $\kappa$ and $r_N$ control the precision with which conformal blocks are computed, and $\ell_{\text{set}}$ is the set of spins included in our sum rules. Higher $\Lambda$ translates to more constraining results.

With regards to \texttt{SDPB}, we use \texttt{Set 1}: \{precision=1024, initialMatrixScalePrimal=1e+20, initialMatrixScaleDual=1e+20, maxComplementarity=1e+70, detectPrimalFeasibleJump, detectDualFeasibleJump \} and \texttt{Set 2}: \{precision=768, initialMatrixScalePrimal=1e+40, initialMatrixScaleDual=1e+40, maxComplementarity=1e+200, detectPrimalFeasibleJump, detectDualFeasibleJump\}.

\end{appendices}

\bibliography{main}

\ifx\mcitethebibliography\mciteundefinedmacro
  \let\mcitethebibliography\thebibliography
  \expandafter\let\csname endmcitethebibliography\endcsname\endthebibliography
  \def\mcitedefaultmidpunct{,~}
  \def\mcitedefaultendpunct{.}
  \def\mcitedefaultseppunct{;}
  \def\EndOfBibitem{}
  \def\mciteBstWouldAddEndPuncttrue{}
  \def\mciteBstWouldAddEndPunctfalse{}
  \def\mciteSetBstMidEndSepPunct#1#2#3{#2}
\fi
\begin{mcitethebibliography}{10}
\ifx\href\asklfhas\newcommand{\href}[2]{#2}\fi
\ifx\arxivref\asklfhas\newcommand{\arxivref}[2]{\href{https://arxiv.org/abs/#1}{#2}}\fi
\ifx\doiref\asklfhas\newcommand{\doiref}[2]{\href{https://doi.org/#1}{#2}}\fi
\parskip 0pt
\normalsize

\bibitem{Rattazzi:2008pe}
R.~Rattazzi, V.~S. Rychkov, E.~Tonni \& A.~Vichi,
\textit{``{Bounding scalar operator dimensions in 4D CFT}''},
\doiref{10.1088/1126-6708/2008/12/031}{JHEP \textbf{0812}, 031 (2008)\ignorespaces}\ignorespaces,
\texttt{\arxivref{0807.0004}{arXiv:0807.0004 \![hep-th]}}\ignorespaces\mciteBstWouldAddEndPuncttrue
\mciteSetBstMidEndSepPunct{\mcitedefaultmidpunct\newline}
{\mcitedefaultendpunct}{\mcitedefaultseppunct}\relax
\EndOfBibitem\bibitem{Poland:2018epd}
D.~Poland, S.~Rychkov \& A.~Vichi,
\textit{``{The Conformal Bootstrap: Theory, Numerical Techniques, and Applications}''},
\doiref{10.1103/RevModPhys.91.015002}{Rev. Mod. Phys. \textbf{91}, 015002 (2019)\ignorespaces}\ignorespaces,
\texttt{\arxivref{1805.04405}{arXiv:1805.04405 \![hep-th]}}\ignorespaces\mciteBstWouldAddEndPuncttrue
\mciteSetBstMidEndSepPunct{\mcitedefaultmidpunct\newline}
{\mcitedefaultendpunct}{\mcitedefaultseppunct}\relax
\EndOfBibitem\bibitem{El-Showk:2012cjh}
S.~El-Showk, M.~F. Paulos, D.~Poland, S.~Rychkov, D.~Simmons-Duffin \& A.~Vichi,
\textit{``{Solving the 3D Ising Model with the Conformal Bootstrap}''},
\doiref{10.1103/PhysRevD.86.025022}{Phys. Rev. D \textbf{86}, 025022 (2012)\ignorespaces}\ignorespaces,
\texttt{\arxivref{1203.6064}{arXiv:1203.6064 \![hep-th]}}\ignorespaces\mciteBstWouldAddEndPuncttrue
\mciteSetBstMidEndSepPunct{\mcitedefaultmidpunct\newline}
{\mcitedefaultendpunct}{\mcitedefaultseppunct}\relax
\EndOfBibitem\bibitem{El-Showk:2014dwa}
S.~El-Showk, M.~F. Paulos, D.~Poland, S.~Rychkov, D.~Simmons-Duffin \& A.~Vichi,
\textit{``{Solving the 3d Ising Model with the Conformal Bootstrap II. c-Minimization and Precise Critical Exponents}''},
\doiref{10.1007/s10955-014-1042-7}{J. Stat. Phys. \textbf{157}, 869 (2014)\ignorespaces}\ignorespaces,
\texttt{\arxivref{1403.4545}{arXiv:1403.4545 \![hep-th]}}\ignorespaces\mciteBstWouldAddEndPuncttrue
\mciteSetBstMidEndSepPunct{\mcitedefaultmidpunct\newline}
{\mcitedefaultendpunct}{\mcitedefaultseppunct}\relax
\EndOfBibitem\bibitem{Kos:2014bka}
F.~Kos, D.~Poland \& D.~Simmons-Duffin,
\textit{``{Bootstrapping Mixed Correlators in the 3D Ising Model}''},
\doiref{10.1007/JHEP11(2014)109}{JHEP \textbf{1411}, 109 (2014)\ignorespaces}\ignorespaces,
\texttt{\arxivref{1406.4858}{arXiv:1406.4858 \![hep-th]}}\ignorespaces\mciteBstWouldAddEndPuncttrue
\mciteSetBstMidEndSepPunct{\mcitedefaultmidpunct\newline}
{\mcitedefaultendpunct}{\mcitedefaultseppunct}\relax
\EndOfBibitem\bibitem{Kos:2016ysd}
F.~Kos, D.~Poland, D.~Simmons-Duffin \& A.~Vichi,
\textit{``{Precision Islands in the Ising and $O(N)$ Models}''},
\doiref{10.1007/JHEP08(2016)036}{JHEP \textbf{1608}, 036 (2016)\ignorespaces}\ignorespaces,
\texttt{\arxivref{1603.04436}{arXiv:1603.04436 \![hep-th]}}\ignorespaces\mciteBstWouldAddEndPuncttrue
\mciteSetBstMidEndSepPunct{\mcitedefaultmidpunct\newline}
{\mcitedefaultendpunct}{\mcitedefaultseppunct}\relax
\EndOfBibitem\bibitem{Chang:2024whx}
C.-H. Chang, V.~Dommes, R.~S. Erramilli, A.~Homrich, P.~Kravchuk, A.~Liu, M.~S. Mitchell, D.~Poland \& D.~Simmons-Duffin,
\textit{``{Bootstrapping the 3d Ising stress tensor}''},
\doiref{10.1007/JHEP03(2025)136}{JHEP \textbf{2503}, 136 (2025)\ignorespaces}\ignorespaces,
\texttt{\arxivref{2411.15300}{arXiv:2411.15300 \![hep-th]}}\ignorespaces\mciteBstWouldAddEndPuncttrue
\mciteSetBstMidEndSepPunct{\mcitedefaultmidpunct\newline}
{\mcitedefaultendpunct}{\mcitedefaultseppunct}\relax
\EndOfBibitem\bibitem{Kos:2013tga}
F.~Kos, D.~Poland \& D.~Simmons-Duffin,
\textit{``{Bootstrapping the $O(N)$ vector models}''},
\doiref{10.1007/JHEP06(2014)091}{JHEP \textbf{1406}, 091 (2014)\ignorespaces}\ignorespaces,
\texttt{\arxivref{1307.6856}{arXiv:1307.6856 \![hep-th]}}\ignorespaces\mciteBstWouldAddEndPuncttrue
\mciteSetBstMidEndSepPunct{\mcitedefaultmidpunct\newline}
{\mcitedefaultendpunct}{\mcitedefaultseppunct}\relax
\EndOfBibitem\bibitem{Kos:2015mba}
F.~Kos, D.~Poland, D.~Simmons-Duffin \& A.~Vichi,
\textit{``{Bootstrapping the O(N) Archipelago}''},
\doiref{10.1007/JHEP11(2015)106}{JHEP \textbf{1511}, 106 (2015)\ignorespaces}\ignorespaces,
\texttt{\arxivref{1504.07997}{arXiv:1504.07997 \![hep-th]}}\ignorespaces\mciteBstWouldAddEndPuncttrue
\mciteSetBstMidEndSepPunct{\mcitedefaultmidpunct\newline}
{\mcitedefaultendpunct}{\mcitedefaultseppunct}\relax
\EndOfBibitem\bibitem{Chester:2019ifh}
S.~M. Chester, W.~Landry, J.~Liu, D.~Poland, D.~Simmons-Duffin, N.~Su \& A.~Vichi,
\textit{``{Carving out OPE space and precise $O(2)$ model critical exponents}''},
\doiref{10.1007/JHEP06(2020)142}{JHEP \textbf{2006}, 142 (2020)\ignorespaces}\ignorespaces,
\texttt{\arxivref{1912.03324}{arXiv:1912.03324 \![hep-th]}}\ignorespaces\mciteBstWouldAddEndPuncttrue
\mciteSetBstMidEndSepPunct{\mcitedefaultmidpunct\newline}
{\mcitedefaultendpunct}{\mcitedefaultseppunct}\relax
\EndOfBibitem\bibitem{Chester:2020iyt}
S.~M. Chester, W.~Landry, J.~Liu, D.~Poland, D.~Simmons-Duffin, N.~Su \& A.~Vichi,
\textit{``{Bootstrapping Heisenberg magnets and their cubic instability}''},
\doiref{10.1103/PhysRevD.104.105013}{Phys. Rev. D \textbf{104}, 105013 (2021)\ignorespaces}\ignorespaces,
\texttt{\arxivref{2011.14647}{arXiv:2011.14647 \![hep-th]}}\ignorespaces\mciteBstWouldAddEndPuncttrue
\mciteSetBstMidEndSepPunct{\mcitedefaultmidpunct\newline}
{\mcitedefaultendpunct}{\mcitedefaultseppunct}\relax
\EndOfBibitem\bibitem{Poland:2011ey}
D.~Poland, D.~Simmons-Duffin \& A.~Vichi,
\textit{``{Carving Out the Space of 4D CFTs}''},
\doiref{10.1007/JHEP05(2012)110}{JHEP \textbf{1205}, 110 (2012)\ignorespaces}\ignorespaces,
\texttt{\arxivref{1109.5176}{arXiv:1109.5176 \![hep-th]}}\ignorespaces\mciteBstWouldAddEndPuncttrue
\mciteSetBstMidEndSepPunct{\mcitedefaultmidpunct\newline}
{\mcitedefaultendpunct}{\mcitedefaultseppunct}\relax
\EndOfBibitem\bibitem{Stergiou:2018gjj}
A.~Stergiou,
\textit{``{Bootstrapping hypercubic and hypertetrahedral theories in three dimensions}''},
\doiref{10.1007/JHEP05(2018)035}{JHEP \textbf{1805}, 035 (2018)\ignorespaces}\ignorespaces,
\texttt{\arxivref{1801.07127}{arXiv:1801.07127 \![hep-th]}}\ignorespaces\mciteBstWouldAddEndPuncttrue
\mciteSetBstMidEndSepPunct{\mcitedefaultmidpunct\newline}
{\mcitedefaultendpunct}{\mcitedefaultseppunct}\relax
\EndOfBibitem\bibitem{Li:2020bnb}
Z.~Li \& D.~Poland,
\textit{``{Searching for gauge theories with the conformal bootstrap}''},
\doiref{10.1007/JHEP03(2021)172}{JHEP \textbf{2103}, 172 (2021)\ignorespaces}\ignorespaces,
\texttt{\arxivref{2005.01721}{arXiv:2005.01721 \![hep-th]}}\ignorespaces\mciteBstWouldAddEndPuncttrue
\mciteSetBstMidEndSepPunct{\mcitedefaultmidpunct\newline}
{\mcitedefaultendpunct}{\mcitedefaultseppunct}\relax
\EndOfBibitem\bibitem{Kravchuk:2021akc}
P.~Kravchuk, D.~Mazac \& S.~Pal,
\textit{``{Automorphic spectra and the conformal bootstrap}''},
\doiref{10.1090/cams/26}{Commun. Am. Math. Soc. \textbf{4}, 1 (2024)\ignorespaces}\ignorespaces,
\texttt{\arxivref{2111.12716}{arXiv:2111.12716 \![hep-th]}}\ignorespaces\mciteBstWouldAddEndPuncttrue
\mciteSetBstMidEndSepPunct{\mcitedefaultmidpunct\newline}
{\mcitedefaultendpunct}{\mcitedefaultseppunct}\relax
\EndOfBibitem\bibitem{Antipin:2019vdg}
O.~Antipin \& J.~Bersini,
\textit{``{Spectrum of anomalous dimensions in hypercubic theories}''},
\doiref{10.1103/PhysRevD.100.065008}{Phys. Rev. D \textbf{100}, 065008 (2019)\ignorespaces}\ignorespaces,
\texttt{\arxivref{1903.04950}{arXiv:1903.04950 \![hep-th]}}\ignorespaces\mciteBstWouldAddEndPuncttrue
\mciteSetBstMidEndSepPunct{\mcitedefaultmidpunct\newline}
{\mcitedefaultendpunct}{\mcitedefaultseppunct}\relax
\EndOfBibitem\bibitem{Bednyakov:2023lfj}
A.~Bednyakov, J.~Henriksson \& S.~R. Kousvos,
\textit{``{Anomalous dimensions in hypercubic theories}''},
\doiref{10.1007/JHEP11(2023)051}{JHEP \textbf{2311}, 051 (2023)\ignorespaces}\ignorespaces,
\texttt{\arxivref{2304.06755}{arXiv:2304.06755 \![hep-th]}}\ignorespaces\mciteBstWouldAddEndPuncttrue
\mciteSetBstMidEndSepPunct{\mcitedefaultmidpunct\newline}
{\mcitedefaultendpunct}{\mcitedefaultseppunct}\relax
\EndOfBibitem\bibitem{Henriksson:2025hwi}
J.~Henriksson, F.~Herzog, S.~R. Kousvos \& J.~Roosmale~Nepveu,
\textit{``{Multi-loop spectra in general scalar EFTs and CFTs}''},
\texttt{\arxivref{2507.12518}{arXiv:2507.12518 \![hep-ph]}}\ignorespaces\mciteBstWouldAddEndPuncttrue
\mciteSetBstMidEndSepPunct{\mcitedefaultmidpunct\newline}
{\mcitedefaultendpunct}{\mcitedefaultseppunct}\relax
\EndOfBibitem\bibitem{Henriksson:2025vyi}
J.~Henriksson, S.~R. Kousvos \& J.~Roosmale~Nepveu,
\textit{``{EFT meets CFT: Multiloop renormalization of higher-dimensional operators in general $\phi^4$ theories}''},
\texttt{\arxivref{2511.16740}{arXiv:2511.16740 \![hep-th]}}\ignorespaces\mciteBstWouldAddEndPuncttrue
\mciteSetBstMidEndSepPunct{\mcitedefaultmidpunct\newline}
{\mcitedefaultendpunct}{\mcitedefaultseppunct}\relax
\EndOfBibitem\bibitem{Brown:1979pq}
L.~S. Brown,
\textit{``{Dimensional Regularization of Composite Operators in Scalar Field Theory}''},
\doiref{10.1016/0003-4916(80)90377-2}{Annals Phys. \textbf{126}, 135 (1980)\ignorespaces}\ignorespaces\mciteBstWouldAddEndPuncttrue
\mciteSetBstMidEndSepPunct{\mcitedefaultmidpunct\newline}
{\mcitedefaultendpunct}{\mcitedefaultseppunct}\relax
\EndOfBibitem\bibitem{Brown:1980qq}
L.~S. Brown \& J.~C. Collins,
\textit{``{Dimensional Renormalization of Scalar Field Theory in Curved Space-time}''},
\doiref{10.1016/0003-4916(80)90232-8}{Annals Phys. \textbf{130}, 215 (1980)\ignorespaces}\ignorespaces\mciteBstWouldAddEndPuncttrue
\mciteSetBstMidEndSepPunct{\mcitedefaultmidpunct\newline}
{\mcitedefaultendpunct}{\mcitedefaultseppunct}\relax
\EndOfBibitem\bibitem{PhysRevB.32.5851}
G.~Murthy \& R.~Shankar,
\textit{``Redundant operators for Ising spins''},
\doiref{10.1103/PhysRevB.32.5851}{Phys. Rev. B \textbf{32}, 5851 (1985)\ignorespaces}\ignorespaces\mciteBstWouldAddEndPuncttrue
\mciteSetBstMidEndSepPunct{\mcitedefaultmidpunct\newline}
{\mcitedefaultendpunct}{\mcitedefaultseppunct}\relax
\EndOfBibitem\bibitem{Rychkov:2015naa}
S.~Rychkov \& Z.~M. Tan,
\textit{``{The $\varepsilon$-expansion from conformal field theory}''},
\doiref{10.1088/1751-8113/48/29/29FT01}{J. Phys. A \textbf{48}, 29FT01 (2015)\ignorespaces}\ignorespaces,
\texttt{\arxivref{1505.00963}{arXiv:1505.00963 \![hep-th]}}\ignorespaces\mciteBstWouldAddEndPuncttrue
\mciteSetBstMidEndSepPunct{\mcitedefaultmidpunct\newline}
{\mcitedefaultendpunct}{\mcitedefaultseppunct}\relax
\EndOfBibitem\bibitem{Pelissetto:2000ek}
A.~Pelissetto \& E.~Vicari,
\textit{``{Critical phenomena and renormalization group theory}''},
\doiref{10.1016/S0370-1573(02)00219-3}{Phys. Rept. \textbf{368}, 549 (2002)\ignorespaces}\ignorespaces,
\texttt{\arxivref{cond-mat/0012164}{cond-mat/0012164}}\ignorespaces\mciteBstWouldAddEndPuncttrue
\mciteSetBstMidEndSepPunct{\mcitedefaultmidpunct\newline}
{\mcitedefaultendpunct}{\mcitedefaultseppunct}\relax
\EndOfBibitem\bibitem{Osborn:2017ucf}
H.~Osborn \& A.~Stergiou,
\textit{``{Seeking fixed points in multiple coupling scalar theories in the $\varepsilon$ expansion}''},
\doiref{10.1007/JHEP05(2018)051}{JHEP \textbf{1805}, 051 (2018)\ignorespaces}\ignorespaces,
\texttt{\arxivref{1707.06165}{arXiv:1707.06165 \![hep-th]}}\ignorespaces\mciteBstWouldAddEndPuncttrue
\mciteSetBstMidEndSepPunct{\mcitedefaultmidpunct\newline}
{\mcitedefaultendpunct}{\mcitedefaultseppunct}\relax
\EndOfBibitem\bibitem{Rychkov:2018vya}
S.~Rychkov \& A.~Stergiou,
\textit{``{General Properties of Multiscalar RG Flows in $d=4-\varepsilon$}''},
\doiref{10.21468/SciPostPhys.6.1.008}{SciPost Phys. \textbf{6}, 008 (2019)\ignorespaces}\ignorespaces,
\texttt{\arxivref{1810.10541}{arXiv:1810.10541 \![hep-th]}}\ignorespaces\mciteBstWouldAddEndPuncttrue
\mciteSetBstMidEndSepPunct{\mcitedefaultmidpunct\newline}
{\mcitedefaultendpunct}{\mcitedefaultseppunct}\relax
\EndOfBibitem\bibitem{PhysRevB.8.3323}
A.~Aharony \& M.~E. Fisher,
\textit{``Critical Behavior of Magnets with Dipolar Interactions. I. Renormalization Group near Four Dimensions''},
\doiref{10.1103/PhysRevB.8.3323}{Phys. Rev. B \textbf{8}, 3323 (1973)\ignorespaces}\ignorespaces\mciteBstWouldAddEndPuncttrue
\mciteSetBstMidEndSepPunct{\mcitedefaultmidpunct\newline}
{\mcitedefaultendpunct}{\mcitedefaultseppunct}\relax
\EndOfBibitem\bibitem{PhysRevB.8.4270}
A.~Aharony,
\textit{``Critical Behavior of Anisotropic Cubic Systems''},
\doiref{10.1103/PhysRevB.8.4270}{Phys. Rev. B \textbf{8}, 4270 (1973)\ignorespaces}\ignorespaces\mciteBstWouldAddEndPuncttrue
\mciteSetBstMidEndSepPunct{\mcitedefaultmidpunct\newline}
{\mcitedefaultendpunct}{\mcitedefaultseppunct}\relax
\EndOfBibitem\bibitem{D_J_Wallace_1973}
D.~J. Wallace,
\textit{``Critical behaviour of anisotropic cubic systems''},
\doiref{10.1088/0022-3719/6/8/007}{Journal of Physics C: Solid State Physics \textbf{6}, 1390 (1973)\ignorespaces}\ignorespaces\mciteBstWouldAddEndPuncttrue
\mciteSetBstMidEndSepPunct{\mcitedefaultmidpunct\newline}
{\mcitedefaultendpunct}{\mcitedefaultseppunct}\relax
\EndOfBibitem\bibitem{Mudrov:1998hi}
A.~I. Mudrov \& K.~B. Varnashev,
\textit{``{New approach to Borel summation of divergent series and critical exponent estimates for an N vector cubic model in three-dimensions from five loop epsilon expansions}''},
\doiref{10.1103/PhysRevE.58.5371}{Phys. Rev. E \textbf{58}, 5371 (1998)\ignorespaces}\ignorespaces,
\texttt{\arxivref{cond-mat/9805081}{cond-mat/9805081}}\ignorespaces\mciteBstWouldAddEndPuncttrue
\mciteSetBstMidEndSepPunct{\mcitedefaultmidpunct\newline}
{\mcitedefaultendpunct}{\mcitedefaultseppunct}\relax
\EndOfBibitem\bibitem{Varnashev:1999ze}
K.~B. Varnashev,
\textit{``{Stability of a cubic fixed point in three-dimensions: Critical exponents for generic N}''},
\doiref{10.1103/PhysRevB.61.14660}{Phys. Rev. B \textbf{61}, 14660 (2000)\ignorespaces}\ignorespaces,
\texttt{\arxivref{cond-mat/9909087}{cond-mat/9909087}}\ignorespaces\mciteBstWouldAddEndPuncttrue
\mciteSetBstMidEndSepPunct{\mcitedefaultmidpunct\newline}
{\mcitedefaultendpunct}{\mcitedefaultseppunct}\relax
\EndOfBibitem\bibitem{Carmona:1999rm}
J.~M. Carmona, A.~Pelissetto \& E.~Vicari,
\textit{``{The N component Ginzburg-Landau Hamiltonian with cubic anisotropy: A Six loop study}''},
\doiref{10.1103/PhysRevB.61.15136}{Phys. Rev. B \textbf{61}, 15136 (2000)\ignorespaces}\ignorespaces,
\texttt{\arxivref{cond-mat/9912115}{cond-mat/9912115}}\ignorespaces\mciteBstWouldAddEndPuncttrue
\mciteSetBstMidEndSepPunct{\mcitedefaultmidpunct\newline}
{\mcitedefaultendpunct}{\mcitedefaultseppunct}\relax
\EndOfBibitem\bibitem{Folk_2000}
R.~Folk, Y.~Holovatch \& T.~Yavors’kii,
\textit{``Pseudo $\varepsilon$ expansion of six-loop renormalization-group functions of an anisotropic cubic model''},
\doiref{10.1103/physrevb.62.12195}{Physical Review B \textbf{62}, 12195–12200 (2000)\ignorespaces}\ignorespaces,
\texttt{\arxivref{cond-mat/0003216}{cond-mat/0003216}}\ignorespaces\mciteBstWouldAddEndPuncttrue
\mciteSetBstMidEndSepPunct{\mcitedefaultmidpunct\newline}
{\mcitedefaultendpunct}{\mcitedefaultseppunct}\relax
\EndOfBibitem\bibitem{Adzhemyan:2019gvv}
L.~T. Adzhemyan, E.~V. Ivanova, M.~V. Kompaniets, A.~Kudlis \& A.~I. Sokolov,
\textit{``{Six-loop $\varepsilon$ expansion study of three-dimensional $n$-vector model with cubic anisotropy}''},
\doiref{10.1016/j.nuclphysb.2019.02.001}{Nucl. Phys. B \textbf{940}, 332 (2019)\ignorespaces}\ignorespaces,
\texttt{\arxivref{1901.02754}{arXiv:1901.02754 \![cond-mat.stat-mech]}}\ignorespaces\mciteBstWouldAddEndPuncttrue
\mciteSetBstMidEndSepPunct{\mcitedefaultmidpunct\newline}
{\mcitedefaultendpunct}{\mcitedefaultseppunct}\relax
\EndOfBibitem\bibitem{Binder:2021vep}
D.~J. Binder,
\textit{``{The cubic fixed point at large $N$}''},
\doiref{10.1007/JHEP09(2021)071}{JHEP \textbf{2109}, 071 (2021)\ignorespaces}\ignorespaces,
\texttt{\arxivref{2106.03493}{arXiv:2106.03493 \![hep-th]}}\ignorespaces\mciteBstWouldAddEndPuncttrue
\mciteSetBstMidEndSepPunct{\mcitedefaultmidpunct\newline}
{\mcitedefaultendpunct}{\mcitedefaultseppunct}\relax
\EndOfBibitem\bibitem{Aharony:2022ajv}
A.~Aharony, O.~Entin-Wohlman \& A.~Kudlis,
\textit{``{Different critical behaviors in cubic to trigonal and tetragonal perovskites}''},
\doiref{10.1103/PhysRevB.105.104101}{Phys. Rev. B \textbf{105}, 104101 (2022)\ignorespaces}\ignorespaces,
\texttt{\arxivref{2201.08252}{arXiv:2201.08252 \![cond-mat.stat-mech]}}\ignorespaces\mciteBstWouldAddEndPuncttrue
\mciteSetBstMidEndSepPunct{\mcitedefaultmidpunct\newline}
{\mcitedefaultendpunct}{\mcitedefaultseppunct}\relax
\EndOfBibitem\bibitem{Hasenbusch:2022zur}
M.~Hasenbusch,
\textit{``{Cubic fixed point in three dimensions: Monte Carlo simulations of the \ensuremath{\phi}4 model on the simple cubic lattice}''},
\doiref{10.1103/PhysRevB.107.024409}{Phys. Rev. B \textbf{107}, 024409 (2023)\ignorespaces}\ignorespaces,
\texttt{\arxivref{2211.16170}{arXiv:2211.16170 \![cond-mat.stat-mech]}}\ignorespaces\mciteBstWouldAddEndPuncttrue
\mciteSetBstMidEndSepPunct{\mcitedefaultmidpunct\newline}
{\mcitedefaultendpunct}{\mcitedefaultseppunct}\relax
\EndOfBibitem\bibitem{Rong:2023owx}
J.~Rong \& N.~Su,
\textit{``{From O(3) to Cubic CFT: Conformal Perturbation and the Large Charge Sector}''},
\texttt{\arxivref{2311.00933}{arXiv:2311.00933 \![hep-th]}}\ignorespaces\mciteBstWouldAddEndPuncttrue
\mciteSetBstMidEndSepPunct{\mcitedefaultmidpunct\newline}
{\mcitedefaultendpunct}{\mcitedefaultseppunct}\relax
\EndOfBibitem\bibitem{Rong:2017cow}
J.~Rong \& N.~Su,
\textit{``{Scalar CFTs and Their Large N Limits}''},
\doiref{10.1007/JHEP09(2018)103}{JHEP \textbf{1809}, 103 (2018)\ignorespaces}\ignorespaces,
\texttt{\arxivref{1712.00985}{arXiv:1712.00985 \![hep-th]}}\ignorespaces\mciteBstWouldAddEndPuncttrue
\mciteSetBstMidEndSepPunct{\mcitedefaultmidpunct\newline}
{\mcitedefaultendpunct}{\mcitedefaultseppunct}\relax
\EndOfBibitem\bibitem{Kousvos:2018rhl}
S.~R. Kousvos \& A.~Stergiou,
\textit{``{Bootstrapping Mixed Correlators in Three-Dimensional Cubic Theories}''},
\doiref{10.21468/SciPostPhys.6.3.035}{SciPost Phys. \textbf{6}, 035 (2019)\ignorespaces}\ignorespaces,
\texttt{\arxivref{1810.10015}{arXiv:1810.10015 \![hep-th]}}\ignorespaces\mciteBstWouldAddEndPuncttrue
\mciteSetBstMidEndSepPunct{\mcitedefaultmidpunct\newline}
{\mcitedefaultendpunct}{\mcitedefaultseppunct}\relax
\EndOfBibitem\bibitem{Kousvos:2019hgc}
S.~R. Kousvos \& A.~Stergiou,
\textit{``{Bootstrapping Mixed Correlators in Three-Dimensional Cubic Theories II}''},
\doiref{10.21468/SciPostPhys.8.6.085}{SciPost Phys. \textbf{8}, 085 (2020)\ignorespaces}\ignorespaces,
\texttt{\arxivref{1911.00522}{arXiv:1911.00522 \![hep-th]}}\ignorespaces\mciteBstWouldAddEndPuncttrue
\mciteSetBstMidEndSepPunct{\mcitedefaultmidpunct\newline}
{\mcitedefaultendpunct}{\mcitedefaultseppunct}\relax
\EndOfBibitem\bibitem{Brezin:1974zr}
E.~Brezin, C.~De~Dominicis \& J.~Zinn-Justin,
\textit{``{Anomalous dimensions of higher-order operators in the $\varphi^4$-theory}''},
\doiref{10.1007/BF02819916}{Lett. Nuovo Cim. \textbf{9S2}, 483 (1974)\ignorespaces}\ignorespaces\mciteBstWouldAddEndPuncttrue
\mciteSetBstMidEndSepPunct{\mcitedefaultmidpunct\newline}
{\mcitedefaultendpunct}{\mcitedefaultseppunct}\relax
\EndOfBibitem\bibitem{Skvortsov:2015pea}
E.~D. Skvortsov,
\textit{``{On (un)broken higher-spin symmetry in vector models.}''},
\texttt{\arxivref{1512.05994}{arXiv:1512.05994 \![hep-th]}}\ignorespaces\mciteBstWouldAddEndPuncttrue
\mciteSetBstMidEndSepPunct{\mcitedefaultmidpunct\newline}
{\mcitedefaultendpunct}{\mcitedefaultseppunct}\relax
\EndOfBibitem\bibitem{Kousvos:2021rar}
S.~R. Kousvos \& A.~Stergiou,
\textit{``{Bootstrapping mixed MN correlators in 3D}''},
\doiref{10.21468/SciPostPhys.12.6.206}{SciPost Phys. \textbf{12}, 206 (2022)\ignorespaces}\ignorespaces,
\texttt{\arxivref{2112.03919}{arXiv:2112.03919 \![hep-th]}}\ignorespaces\mciteBstWouldAddEndPuncttrue
\mciteSetBstMidEndSepPunct{\mcitedefaultmidpunct\newline}
{\mcitedefaultendpunct}{\mcitedefaultseppunct}\relax
\EndOfBibitem\bibitem{Kousvos:2024dlz}
S.~R. Kousvos, A.~Piazza \& A.~Vichi,
\textit{``{Exploring replica-Potts CFTs in two dimensions}''},
\doiref{10.1007/JHEP11(2024)030}{JHEP \textbf{2411}, 030 (2024)\ignorespaces}\ignorespaces,
\texttt{\arxivref{2405.19416}{arXiv:2405.19416 \![hep-th]}}\ignorespaces\mciteBstWouldAddEndPuncttrue
\mciteSetBstMidEndSepPunct{\mcitedefaultmidpunct\newline}
{\mcitedefaultendpunct}{\mcitedefaultseppunct}\relax
\EndOfBibitem\bibitem{Bednyakov:2021ojn}
A.~Bednyakov \& A.~Pikelner,
\textit{``{Six-loop beta functions in general scalar theory}''},
\doiref{10.1007/JHEP04(2021)233}{JHEP \textbf{2104}, 233 (2021)\ignorespaces}\ignorespaces,
\texttt{\arxivref{2102.12832}{arXiv:2102.12832 \![hep-ph]}}\ignorespaces\mciteBstWouldAddEndPuncttrue
\mciteSetBstMidEndSepPunct{\mcitedefaultmidpunct\newline}
{\mcitedefaultendpunct}{\mcitedefaultseppunct}\relax
\EndOfBibitem\bibitem{Henriksson:2022rnm}
J.~Henriksson,
\textit{``{The critical O(N) CFT: Methods and conformal data}''},
\doiref{10.1016/j.physrep.2022.12.002}{Phys. Rept. \textbf{1002}, 1 (2023)\ignorespaces}\ignorespaces,
\texttt{\arxivref{2201.09520}{arXiv:2201.09520 \![hep-th]}}\ignorespaces\mciteBstWouldAddEndPuncttrue
\mciteSetBstMidEndSepPunct{\mcitedefaultmidpunct\newline}
{\mcitedefaultendpunct}{\mcitedefaultseppunct}\relax
\EndOfBibitem\bibitem{Hasenbusch:2011zwv}
M.~Hasenbusch \& E.~Vicari,
\textit{``{Anisotropic perturbations in three-dimensional O(N)-symmetric vector models}''},
\doiref{10.1103/PhysRevB.84.125136}{Phys. Rev. B \textbf{84}, 125136 (2011)\ignorespaces}\ignorespaces\mciteBstWouldAddEndPuncttrue
\mciteSetBstMidEndSepPunct{\mcitedefaultmidpunct\newline}
{\mcitedefaultendpunct}{\mcitedefaultseppunct}\relax
\EndOfBibitem\bibitem{Stergiou:2019dcv}
A.~Stergiou,
\textit{``{Bootstrapping MN and Tetragonal CFTs in Three Dimensions}''},
\doiref{10.21468/SciPostPhys.7.1.010}{SciPost Phys. \textbf{7}, 010 (2019)\ignorespaces}\ignorespaces,
\texttt{\arxivref{1904.00017}{arXiv:1904.00017 \![hep-th]}}\ignorespaces\mciteBstWouldAddEndPuncttrue
\mciteSetBstMidEndSepPunct{\mcitedefaultmidpunct\newline}
{\mcitedefaultendpunct}{\mcitedefaultseppunct}\relax
\EndOfBibitem\bibitem{Henriksson:2021lwn}
J.~Henriksson \& A.~Stergiou,
\textit{``{Perturbative and Nonperturbative Studies of CFTs with MN Global Symmetry}''},
\doiref{10.21468/SciPostPhys.11.1.015}{SciPost Phys. \textbf{11}, 015 (2021)\ignorespaces}\ignorespaces,
\texttt{\arxivref{2101.08788}{arXiv:2101.08788 \![hep-th]}}\ignorespaces\mciteBstWouldAddEndPuncttrue
\mciteSetBstMidEndSepPunct{\mcitedefaultmidpunct\newline}
{\mcitedefaultendpunct}{\mcitedefaultseppunct}\relax
\EndOfBibitem\bibitem{Nakayama:2014lva}
Y.~Nakayama \& T.~Ohtsuki,
\textit{``{Approaching the conformal window of $O(n)\times O(m)$ symmetric Landau-Ginzburg models using the conformal bootstrap}''},
\doiref{10.1103/PhysRevD.89.126009}{Phys. Rev. D \textbf{89}, 126009 (2014)\ignorespaces}\ignorespaces,
\texttt{\arxivref{1404.0489}{arXiv:1404.0489 \![hep-th]}}\ignorespaces\mciteBstWouldAddEndPuncttrue
\mciteSetBstMidEndSepPunct{\mcitedefaultmidpunct\newline}
{\mcitedefaultendpunct}{\mcitedefaultseppunct}\relax
\EndOfBibitem\bibitem{Nakayama:2014sba}
Y.~Nakayama \& T.~Ohtsuki,
\textit{``{Bootstrapping phase transitions in QCD and frustrated spin systems}''},
\doiref{10.1103/PhysRevD.91.021901}{Phys. Rev. D \textbf{91}, 021901 (2015)\ignorespaces}\ignorespaces,
\texttt{\arxivref{1407.6195}{arXiv:1407.6195 \![hep-th]}}\ignorespaces\mciteBstWouldAddEndPuncttrue
\mciteSetBstMidEndSepPunct{\mcitedefaultmidpunct\newline}
{\mcitedefaultendpunct}{\mcitedefaultseppunct}\relax
\EndOfBibitem\bibitem{Henriksson:2020fqi}
J.~Henriksson, S.~R. Kousvos \& A.~Stergiou,
\textit{``{Analytic and Numerical Bootstrap of CFTs with $O(m)\times O(n)$ Global Symmetry in 3D}''},
\doiref{10.21468/SciPostPhys.9.3.035}{SciPost Phys. \textbf{9}, 035 (2020)\ignorespaces}\ignorespaces,
\texttt{\arxivref{2004.14388}{arXiv:2004.14388 \![hep-th]}}\ignorespaces\mciteBstWouldAddEndPuncttrue
\mciteSetBstMidEndSepPunct{\mcitedefaultmidpunct\newline}
{\mcitedefaultendpunct}{\mcitedefaultseppunct}\relax
\EndOfBibitem\bibitem{Dowens:2020cua}
M.~T. Dowens \& C.~A. Hooley,
\textit{``{Multi-fixed point numerical conformal bootstrap: a case study with structured global symmetry}''},
\doiref{10.1007/JHEP03(2021)147}{JHEP \textbf{2021}, 147 (2020)\ignorespaces}\ignorespaces,
\texttt{\arxivref{2004.14978}{arXiv:2004.14978 \![hep-th]}}\ignorespaces\mciteBstWouldAddEndPuncttrue
\mciteSetBstMidEndSepPunct{\mcitedefaultmidpunct\newline}
{\mcitedefaultendpunct}{\mcitedefaultseppunct}\relax
\EndOfBibitem\bibitem{Reehorst:2024vyq}
M.~Reehorst, S.~Rychkov, B.~Sirois \& B.~C. van~Rees,
\textit{``{Bootstrapping frustrated magnets: the fate of the chiral ${\rm O}(N)\times {\rm O}(2)$ universality class}''},
\doiref{10.21468/SciPostPhys.18.2.060}{SciPost Phys. \textbf{18}, 060 (2025)\ignorespaces}\ignorespaces,
\texttt{\arxivref{2405.19411}{arXiv:2405.19411 \![hep-th]}}\ignorespaces\mciteBstWouldAddEndPuncttrue
\mciteSetBstMidEndSepPunct{\mcitedefaultmidpunct\newline}
{\mcitedefaultendpunct}{\mcitedefaultseppunct}\relax
\EndOfBibitem\bibitem{Osborn:2020cnf}
H.~Osborn \& A.~Stergiou,
\textit{``{Heavy handed quest for fixed points in multiple coupling scalar theories in the $\varepsilon$ expansion}''},
\doiref{10.1007/JHEP04(2021)128}{JHEP \textbf{2104}, 128 (2021)\ignorespaces}\ignorespaces,
\texttt{\arxivref{2010.15915}{arXiv:2010.15915 \![hep-th]}}\ignorespaces\mciteBstWouldAddEndPuncttrue
\mciteSetBstMidEndSepPunct{\mcitedefaultmidpunct\newline}
{\mcitedefaultendpunct}{\mcitedefaultseppunct}\relax
\EndOfBibitem\bibitem{Nelson:1974xnq}
D.~R. Nelson, J.~M. Kosterlitz \& M.~E. Fisher,
\textit{``{Renormalization-Group Analysis of Bicritical and Tetracritical Points}''},
\doiref{10.1103/PhysRevLett.33.813}{Phys. Rev. Lett. \textbf{33}, 813 (1974)\ignorespaces}\ignorespaces\mciteBstWouldAddEndPuncttrue
\mciteSetBstMidEndSepPunct{\mcitedefaultmidpunct\newline}
{\mcitedefaultendpunct}{\mcitedefaultseppunct}\relax
\EndOfBibitem\bibitem{Kosterlitz:1976zza}
J.~M. Kosterlitz, D.~R. Nelson \& M.~E. Fisher,
\textit{``{Bicritical and tetracritical points in anisotropic antiferromagnetic systems}''},
\doiref{10.1103/PhysRevB.13.412}{Phys. Rev. B \textbf{13}, 412 (1976)\ignorespaces}\ignorespaces\mciteBstWouldAddEndPuncttrue
\mciteSetBstMidEndSepPunct{\mcitedefaultmidpunct\newline}
{\mcitedefaultendpunct}{\mcitedefaultseppunct}\relax
\EndOfBibitem\bibitem{Calabrese:2002bm}
P.~Calabrese, A.~Pelissetto \& E.~Vicari,
\textit{``{Multicritical phenomena in O(n(1)) + O(n(2)) symmetric theories}''},
\doiref{10.1103/PhysRevB.67.054505}{Phys. Rev. B \textbf{67}, 054505 (2003)\ignorespaces}\ignorespaces,
\texttt{\arxivref{cond-mat/0209580}{cond-mat/0209580}}\ignorespaces\mciteBstWouldAddEndPuncttrue
\mciteSetBstMidEndSepPunct{\mcitedefaultmidpunct\newline}
{\mcitedefaultendpunct}{\mcitedefaultseppunct}\relax
\EndOfBibitem\bibitem{Benedetti:2020rrq}
D.~Benedetti, R.~Gurau, S.~Harribey \& K.~Suzuki,
\textit{``{Long-range multi-scalar models at three loops}''},
\doiref{10.1088/1751-8121/abb6ae}{J. Phys. A \textbf{53}, 445008 (2020)\ignorespaces}\ignorespaces,
\texttt{\arxivref{2007.04603}{arXiv:2007.04603 \![hep-th]}}\ignorespaces\mciteBstWouldAddEndPuncttrue
\mciteSetBstMidEndSepPunct{\mcitedefaultmidpunct\newline}
{\mcitedefaultendpunct}{\mcitedefaultseppunct}\relax
\EndOfBibitem\bibitem{Liendo:2022bmv}
P.~Liendo, J.~Rong \& H.~Zhang,
\textit{``{Spontaneous breaking of finite group symmetries at all temperatures}''},
\doiref{10.21468/SciPostPhys.14.6.168}{SciPost Phys. \textbf{14}, 168 (2023)\ignorespaces}\ignorespaces,
\texttt{\arxivref{2205.13964}{arXiv:2205.13964 \![hep-th]}}\ignorespaces\mciteBstWouldAddEndPuncttrue
\mciteSetBstMidEndSepPunct{\mcitedefaultmidpunct\newline}
{\mcitedefaultendpunct}{\mcitedefaultseppunct}\relax
\EndOfBibitem\bibitem{Go:2019lke}
M.~Go \& Y.~Tachikawa,
\textit{``{autoboot: A generator of bootstrap equations with global symmetry}''},
\doiref{10.1007/JHEP06(2019)084}{JHEP \textbf{1906}, 084 (2019)\ignorespaces}\ignorespaces,
\texttt{\arxivref{1903.10522}{arXiv:1903.10522 \![hep-th]}}\ignorespaces\mciteBstWouldAddEndPuncttrue
\mciteSetBstMidEndSepPunct{\mcitedefaultmidpunct\newline}
{\mcitedefaultendpunct}{\mcitedefaultseppunct}\relax
\EndOfBibitem\bibitem{Go:2020ahx}
M.~Go,
\textit{``{An Automated Generation of Bootstrap Equations for Numerical Study of Critical Phenomena}''},
\texttt{\arxivref{2006.04173}{arXiv:2006.04173 \![hep-th]}}\ignorespaces\mciteBstWouldAddEndPuncttrue
\mciteSetBstMidEndSepPunct{\mcitedefaultmidpunct\newline}
{\mcitedefaultendpunct}{\mcitedefaultseppunct}\relax
\EndOfBibitem\bibitem{Simpleboot}
N.~Su,
\textit{``{Simpleboot}''},
\href{https://gitlab.com/bootstrapcollaboration/simpleboot}{\nolinkurl{https://gitlab.com/bootstrapcollaboration/simpleboot}}\mciteBstWouldAddEndPuncttrue
\mciteSetBstMidEndSepPunct{\mcitedefaultmidpunct\newline}
{\mcitedefaultendpunct}{\mcitedefaultseppunct}\relax
\EndOfBibitem\bibitem{Simmons-Duffin:2015qma}
D.~Simmons-Duffin,
\textit{``{A Semidefinite Program Solver for the Conformal Bootstrap}''},
\doiref{10.1007/JHEP06(2015)174}{JHEP \textbf{1506}, 174 (2015)\ignorespaces}\ignorespaces,
\texttt{\arxivref{1502.02033}{arXiv:1502.02033 \![hep-th]}}\ignorespaces\mciteBstWouldAddEndPuncttrue
\mciteSetBstMidEndSepPunct{\mcitedefaultmidpunct\newline}
{\mcitedefaultendpunct}{\mcitedefaultseppunct}\relax
\EndOfBibitem\bibitem{Landry:2019qug}
W.~Landry \& D.~Simmons-Duffin,
\textit{``{Scaling the semidefinite program solver SDPB}''},
\texttt{\arxivref{1909.09745}{arXiv:1909.09745 \![hep-th]}}\ignorespaces\mciteBstWouldAddEndPuncttrue
\mciteSetBstMidEndSepPunct{\mcitedefaultmidpunct\newline}
{\mcitedefaultendpunct}{\mcitedefaultseppunct}\relax
\EndOfBibitem\bibitem{PerimeterCourse}
A.~Liu \& N.~Su,
\textit{``{Mini-Course of Numerical Conformal Bootstrap}''},
\href{https://events.perimeterinstitute.ca/event/45/}{\nolinkurl{https://events.perimeterinstitute.ca/event/45/}}\mciteBstWouldAddEndPuncttrue
\mciteSetBstMidEndSepPunct{\mcitedefaultmidpunct\newline}
{\mcitedefaultendpunct}{\mcitedefaultseppunct}\relax
\EndOfBibitem\bibitem{PerimeterCourseTutorials}
A.~Liu \& N.~Su,
\textit{``{Mini-Course of Numerical Conformal Bootstrap, Tutorials}''},
\href{https://gitlab.com/AikeLiu/Bootstrap-Mini-Course/-/tree/master}{\nolinkurl{https://gitlab.com/AikeLiu/Bootstrap-Mini-Course/-/tree/master}}\mciteBstWouldAddEndPuncttrue
\mciteSetBstMidEndSepPunct{\mcitedefaultmidpunct\newline}
{\mcitedefaultendpunct}{\mcitedefaultseppunct}\relax
\EndOfBibitem\bibitem{Paulos:2019gtx}
M.~F. Paulos,
\textit{``{Analytic functional bootstrap for CFTs in $d > 1$}''},
\doiref{10.1007/JHEP04(2020)093}{JHEP \textbf{2004}, 093 (2020)\ignorespaces}\ignorespaces,
\texttt{\arxivref{1910.08563}{arXiv:1910.08563 \![hep-th]}}\ignorespaces\mciteBstWouldAddEndPuncttrue
\mciteSetBstMidEndSepPunct{\mcitedefaultmidpunct\newline}
{\mcitedefaultendpunct}{\mcitedefaultseppunct}\relax
\EndOfBibitem\bibitem{Ghosh:2023onl}
K.~Ghosh \& Z.~Zheng,
\textit{``{Numerical conformal bootstrap with analytic functionals and outer approximation}''},
\doiref{10.1007/JHEP09(2024)143}{JHEP \textbf{2409}, 143 (2024)\ignorespaces}\ignorespaces,
\texttt{\arxivref{2307.11144}{arXiv:2307.11144 \![hep-th]}}\ignorespaces\mciteBstWouldAddEndPuncttrue
\mciteSetBstMidEndSepPunct{\mcitedefaultmidpunct\newline}
{\mcitedefaultendpunct}{\mcitedefaultseppunct}\relax
\EndOfBibitem\bibitem{Kousvos:2022ewl}
S.~R. Kousvos \& A.~Stergiou,
\textit{``{CFTs with $U(m)\times U(n)$ global symmetry in 3D and the chiral phase transition of QCD}''},
\doiref{10.21468/SciPostPhys.15.2.075}{SciPost Phys. \textbf{15}, 075 (2023)\ignorespaces}\ignorespaces,
\texttt{\arxivref{2209.02837}{arXiv:2209.02837 \![hep-th]}}\ignorespaces\mciteBstWouldAddEndPuncttrue
\mciteSetBstMidEndSepPunct{\mcitedefaultmidpunct\newline}
{\mcitedefaultendpunct}{\mcitedefaultseppunct}\relax
\EndOfBibitem\bibitem{Rong:2018okz}
J.~Rong \& N.~Su,
\textit{``{Bootstrapping the minimal $ \mathcal{N} $ = 1 superconformal field theory in three dimensions}''},
\doiref{10.1007/JHEP06(2021)154}{JHEP \textbf{2106}, 154 (2021)\ignorespaces}\ignorespaces,
\texttt{\arxivref{1807.04434}{arXiv:1807.04434 \![hep-th]}}\ignorespaces\mciteBstWouldAddEndPuncttrue
\mciteSetBstMidEndSepPunct{\mcitedefaultmidpunct\newline}
{\mcitedefaultendpunct}{\mcitedefaultseppunct}\relax
\EndOfBibitem\bibitem{Atanasov:2018kqw}
A.~Atanasov, A.~Hillman \& D.~Poland,
\textit{``{Bootstrapping the Minimal 3D SCFT}''},
\doiref{10.1007/JHEP11(2018)140}{JHEP \textbf{1811}, 140 (2018)\ignorespaces}\ignorespaces,
\texttt{\arxivref{1807.05702}{arXiv:1807.05702 \![hep-th]}}\ignorespaces\mciteBstWouldAddEndPuncttrue
\mciteSetBstMidEndSepPunct{\mcitedefaultmidpunct\newline}
{\mcitedefaultendpunct}{\mcitedefaultseppunct}\relax
\EndOfBibitem\bibitem{Atanasov:2022bpi}
A.~Atanasov, A.~Hillman, D.~Poland, J.~Rong \& N.~Su,
\textit{``{Precision bootstrap for the $ \mathcal{N} $ = 1 super-Ising model}''},
\doiref{10.1007/JHEP08(2022)136}{JHEP \textbf{2208}, 136 (2022)\ignorespaces}\ignorespaces,
\texttt{\arxivref{2201.02206}{arXiv:2201.02206 \![hep-th]}}\ignorespaces\mciteBstWouldAddEndPuncttrue
\mciteSetBstMidEndSepPunct{\mcitedefaultmidpunct\newline}
{\mcitedefaultendpunct}{\mcitedefaultseppunct}\relax
\EndOfBibitem\bibitem{Erramilli:2022kgp}
R.~S. Erramilli, L.~V. Iliesiu, P.~Kravchuk, A.~Liu, D.~Poland \& D.~Simmons-Duffin,
\textit{``{The Gross-Neveu-Yukawa archipelago}''},
\doiref{10.1007/JHEP02(2023)036}{JHEP \textbf{2302}, 036 (2023)\ignorespaces}\ignorespaces,
\texttt{\arxivref{2210.02492}{arXiv:2210.02492 \![hep-th]}}\ignorespaces\mciteBstWouldAddEndPuncttrue
\mciteSetBstMidEndSepPunct{\mcitedefaultmidpunct\newline}
{\mcitedefaultendpunct}{\mcitedefaultseppunct}\relax
\EndOfBibitem\bibitem{Mitchell:2024hix}
M.~S. Mitchell \& D.~Poland,
\textit{``{Bounding irrelevant operators in the 3d Gross-Neveu-Yukawa CFTs}''},
\doiref{10.1007/JHEP09(2024)134}{JHEP \textbf{2409}, 134 (2024)\ignorespaces}\ignorespaces,
\texttt{\arxivref{2406.12974}{arXiv:2406.12974 \![hep-th]}}\ignorespaces\mciteBstWouldAddEndPuncttrue
\mciteSetBstMidEndSepPunct{\mcitedefaultmidpunct\newline}
{\mcitedefaultendpunct}{\mcitedefaultseppunct}\relax
\EndOfBibitem\bibitem{Pannell:2023tzc}
W.~H. Pannell \& A.~Stergiou,
\textit{``{Scalar-fermion fixed points in the \ensuremath{\varepsilon} expansion}''},
\doiref{10.1007/JHEP08(2023)128}{JHEP \textbf{2308}, 128 (2023)\ignorespaces}\ignorespaces,
\texttt{\arxivref{2305.14417}{arXiv:2305.14417 \![hep-th]}}\ignorespaces\mciteBstWouldAddEndPuncttrue
\mciteSetBstMidEndSepPunct{\mcitedefaultmidpunct\newline}
{\mcitedefaultendpunct}{\mcitedefaultseppunct}\relax
\EndOfBibitem\bibitem{He:2021xvg}
Y.-C. He, J.~Rong \& N.~Su,
\textit{``{A roadmap for bootstrapping critical gauge theories: decoupling operators of conformal field theories in $d>2$ dimensions}''},
\doiref{10.21468/SciPostPhys.11.6.111}{SciPost Phys. \textbf{11}, 111 (2021)\ignorespaces}\ignorespaces,
\texttt{\arxivref{2101.07262}{arXiv:2101.07262 \![hep-th]}}\ignorespaces\mciteBstWouldAddEndPuncttrue
\mciteSetBstMidEndSepPunct{\mcitedefaultmidpunct\newline}
{\mcitedefaultendpunct}{\mcitedefaultseppunct}\relax
\EndOfBibitem\bibitem{Dotsenko:1998gyp}
V.~Dotsenko, J.~L. Jacobsen, M.-A. Lewis \& M.~Picco,
\textit{``{Coupled Potts models: Self-duality and fixed point structure}''},
\doiref{10.1016/S0550-3213(99)00097-8}{Nucl. Phys. B \textbf{546}, 505 (1999)\ignorespaces}\ignorespaces,
\texttt{\arxivref{cond-mat/9812227}{cond-mat/9812227}}\ignorespaces\mciteBstWouldAddEndPuncttrue
\mciteSetBstMidEndSepPunct{\mcitedefaultmidpunct\newline}
{\mcitedefaultendpunct}{\mcitedefaultseppunct}\relax
\EndOfBibitem\bibitem{Antunes:2022vtb}
A.~Antunes \& C.~Behan,
\textit{``{Coupled Minimal Conformal Field Theory Models Revisited}''},
\doiref{10.1103/PhysRevLett.130.071602}{Phys. Rev. Lett. \textbf{130}, 071602 (2023)\ignorespaces}\ignorespaces,
\texttt{\arxivref{2211.16503}{arXiv:2211.16503 \![hep-th]}}\ignorespaces\mciteBstWouldAddEndPuncttrue
\mciteSetBstMidEndSepPunct{\mcitedefaultmidpunct\newline}
{\mcitedefaultendpunct}{\mcitedefaultseppunct}\relax
\EndOfBibitem\bibitem{Antunes:2024mfb}
A.~Antunes \& C.~Behan,
\textit{``{Coupled minimal models revisited II: Constraints from permutation symmetry}''},
\doiref{10.21468/SciPostPhys.18.4.132}{SciPost Phys. \textbf{18}, 132 (2025)\ignorespaces}\ignorespaces,
\texttt{\arxivref{2412.21107}{arXiv:2412.21107 \![hep-th]}}\ignorespaces\mciteBstWouldAddEndPuncttrue
\mciteSetBstMidEndSepPunct{\mcitedefaultmidpunct\newline}
{\mcitedefaultendpunct}{\mcitedefaultseppunct}\relax
\EndOfBibitem\bibitem{Henriksson:2025kws}
J.~Henriksson,
\textit{``{The tricritical Ising CFT and conformal bootstrap}''},
\texttt{\arxivref{2501.18711}{arXiv:2501.18711 \![hep-th]}}\ignorespaces\mciteBstWouldAddEndPuncttrue
\mciteSetBstMidEndSepPunct{\mcitedefaultmidpunct\newline}
{\mcitedefaultendpunct}{\mcitedefaultseppunct}\relax
\EndOfBibitem\bibitem{Reehorst:2021ykw}
M.~Reehorst, S.~Rychkov, D.~Simmons-Duffin, B.~Sirois, N.~Su \& B.~van~Rees,
\textit{``{Navigator Function for the Conformal Bootstrap}''},
\doiref{10.21468/SciPostPhys.11.3.072}{SciPost Phys. \textbf{11}, 072 (2021)\ignorespaces}\ignorespaces,
\texttt{\arxivref{2104.09518}{arXiv:2104.09518 \![hep-th]}}\ignorespaces\mciteBstWouldAddEndPuncttrue
\mciteSetBstMidEndSepPunct{\mcitedefaultmidpunct\newline}
{\mcitedefaultendpunct}{\mcitedefaultseppunct}\relax
\EndOfBibitem\bibitem{Liu:2023elz}
A.~Liu, D.~Simmons-Duffin, N.~Su \& B.~C. van~Rees,
\textit{``{Skydiving to Bootstrap Islands}''},
\texttt{\arxivref{2307.13046}{arXiv:2307.13046 \![hep-th]}}\ignorespaces\mciteBstWouldAddEndPuncttrue
\mciteSetBstMidEndSepPunct{\mcitedefaultmidpunct\newline}
{\mcitedefaultendpunct}{\mcitedefaultseppunct}\relax
\EndOfBibitem\bibitem{CREATE}
{\relax King's College London},
\textit{``{King's Computational Research, Engineering and Technology Environment (CREATE)}''},
\href{https://doi.org/10.18742/rnvf-m076}{\nolinkurl{https://doi.org/10.18742/rnvf-m076}}\mciteBstWouldAddEndPuncttrue
\mciteSetBstMidEndSepPunct{\mcitedefaultmidpunct\newline}
{\mcitedefaultendpunct}{\mcitedefaultseppunct}\relax
\EndOfBibitem\end{mcitethebibliography}

\end{document}